\documentclass[sigconf]{acmart}
\usepackage{listings}
\usepackage{xcolor}
\usepackage{float}

\floatstyle{plain}
\newfloat{listing}{htbp}{lop}
\floatname{listing}{Listing}

\usepackage{amsmath}
\usepackage{mathtools} 
\usepackage{makecell}
\usepackage{subcaption}
\usepackage{multirow}
\usepackage{enumitem}
\usepackage[group-separator={,},group-minimum-digits={3}]{siunitx}

\newif\ifshowextra
\showextrafalse % Using this for the arxiv version

\AtBeginDocument{%
  }

\setcopyright{acmlicensed}
\copyrightyear{2025}
\acmYear{2025}
\setcopyright{cc}
\setcctype{by}
\acmConference[SC '25]{The International Conference for High Performance Computing, Networking, Storage and Analysis}{November 16--21, 2025}{St Louis, MO, USA}
\acmBooktitle{The International Conference for High Performance Computing, Networking, Storage and Analysis (SC '25), November 16--21, 2025, St Louis, MO, USA}
\acmDOI{10.1145/3712285.3759835}
\acmISBN{979-8-4007-1466-5/2025/11}

\usepackage{xspace}

\begin{document}

%%
%% The "title" command has an optional parameter,
%% allowing the author to define a "short title" to be used in page headers.
\title{Bine Trees: Enhancing Collective Operations\\by Optimizing Communication Locality}

%%
%% The "author" command and its associated commands are used to define
%% the authors and their affiliations.
%% Of note is the shared affiliation of the first two authors, and the
%% "authornote" and "authornotemark" commands
%% used to denote shared contribution to the research.
%\author{Anonymous Authors}
\iftrue
\author{Daniele De Sensi}
\affiliation{%
  \institution{Sapienza University of Rome}
  \city{Rome}
  %\state{Ohio}
  \country{Italy}
}
\email{desensi@di.uniroma1.it}

\author{Saverio Pasqualoni}
\affiliation{%
  \institution{Sapienza University of Rome}
  \city{Rome}
  %\state{Ohio}
  \country{Italy}
}
\email{pasqualoni.1845572@studenti.uniroma1.it}

\author{Lorenzo Piarulli}
\affiliation{%
  \institution{Sapienza University of Rome}
  \city{Rome}
  %\state{Ohio}
  \country{Italy}
}
\email{piarulli@di.uniroma1.it}

\author{Tommaso Bonato}
\affiliation{%
  \institution{ETH Zurich}
  \city{Zurich}
  %\state{Ohio}
  \country{Switzerland}
}
\email{tommaso.bonato@inf.ethz.ch}

\author{Seydou Ba}
\affiliation{%
  \institution{RIKEN}
  \city{Kobe}
  %\state{Ohio}
  \country{Japan}
}
\email{seydou.ba@riken.jp}

\author{Matteo Turisini}
\affiliation{%
  \institution{CINECA}
  \city{Rome}
  %\state{Ohio}
  \country{Italy}
}
\email{m.turisini@cineca.it}

\author{Jens Domke}
\affiliation{%
  \institution{RIKEN}
  \city{Kobe}
  %\state{Ohio}
  \country{Japan}
}
\email{jens.domke@riken.jp}

\author{Torsten Hoefler}
\affiliation{%
  \institution{ETH Zurich}
  \city{Zurich}
  %\state{Ohio}
  \country{Switzerland}
}
\email{torsten.hoefler@inf.ethz.ch}
\fi

%%
%% By default, the full list of authors will be used in the page
%% headers. Often, this list is too long, and will overlap
%% other information printed in the page headers. This command allows
%% the author to define a more concise list
%% of authors' names for this purpose.
\renewcommand{\shortauthors}{De Sensi et al.}

\newcommand{\dan}[1]{\textcolor{red}{[Daniele: #1]}}

\newcommand{\bine}{\emph{Bine}\xspace}
\newcommand{\Bine}{\emph{Bine}\xspace}

\newcommand{\bcast}{broadcast\xspace}
\newcommand{\reduce}{reduce\xspace}
\newcommand{\gat}{gather\xspace}
\newcommand{\scatter}{scatter\xspace}
\newcommand{\reducescatter}{reduce-scatter\xspace}
\newcommand{\allgather}{allgather\xspace}
\newcommand{\allreduce}{allreduce\xspace}
\newcommand{\alltoall}{alltoall\xspace}

\newcommand{\leonardo}{Leonardo\xspace}
\newcommand{\fugaku}{Fugaku\xspace}
\newcommand{\lumi}{LUMI\xspace}
\newcommand{\mn}{MareNostrum 5\xspace}

\newcommand{\quotes}[1]{``#1''}

%%
%% The abstract is a short summary of the work to be presented in the
%% article.
\begin{abstract}
  %As high-performance computing (HPC) systems grow, optimizing communication locality becomes essential for performance. HPC networks are often oversubscribed, consisting of fully connected groups that are sparsely connected. We introduce Binomial Negabinary (\Bine) trees, a novel approach to enhance collective operations by reducing inter-group communication. \Bine trees reduce the distance between communicating ranks, thus reducing traffic on global links and alleviating congestion. 
  %Bine trees are topology-agnostic and do not assume any rank partitioning, making them ideal for production supercomputers with irregular allocations. We design algorithms for eight collectives, achieving up to $5\times$ speedups and 33\% less global traffic on four supercomputers with four different topologies. Our results emphasize \bine trees effectiveness in improving performance while reducing the load on global links.
  Communication locality plays a key role in the performance of collective operations on large HPC systems, especially on oversubscribed networks where groups of nodes are fully connected internally but sparsely linked through global connections. We present \Bine (\textit{\underline{bi}nomial \underline{ne}gabinary}) trees, a family of collective algorithms that improve communication locality.
  \Bine trees maintain the generality of binomial trees and butterflies while cutting global-link traffic by up to $33\%$. We implement eight \Bine-based collectives and evaluate them on four large-scale supercomputers with Dragonfly, Dragonfly+, oversubscribed fat-tree, and torus topologies, achieving up to $5\times$ speedups and consistent reductions in global-link traffic across different vector sizes and node counts.
\end{abstract}

%%
%% The code below is generated by the tool at http://dl.acm.org/ccs.cfm.
%% Please copy and paste the code instead of the example below.
%%
\iftrue


%%
%% Keywords. The author(s) should pick words that accurately describe
%% the work being presented. Separate the keywords with commas.
 % TODO
%% A "teaser" image appears between the author and affiliation
%% information and the body of the document, and typically spans the
%% page.

%\received{20 February 2007}
%\received[revised]{12 March 2009}
%\received[accepted]{5 June 2009}
\fi

%%
%% This command processes the author and affiliation and title
%% information and builds the first part of the formatted document.
\maketitle

\section{Introduction}\label{sec:intro}
As high-performance computing (HPC) systems and data centers continue to grow~\cite{top500,10.1145/3581784.3607089,metacluster,colossus}, communication locality plays a crucial role in determining performance. In most HPC networks, communication with nearby nodes is generally faster than with distant nodes. This is particularly evident in scale-up networks, where bandwidth can be up to an order of magnitude higher than that of the scale-out network~\cite{gpugpuinterconnect,hxmesh}. However, even when focusing solely on the scale-out network, locality remains important. While full-bandwidth scale-out networks such as fat trees can deliver high performance, they also come with significant cost and power consumption. As a result, \textit{oversubscribed} (\textit{tapered}) networks are often used. These are typically organized into fully connected \textit{groups}, such as groups in Dragonfly/Dragonfly+ topologies~\cite{dragonfly,dragonflyplus,sc2020,frontier} or subtrees in oversubscribed fat trees, which are then interconnected more sparsely through \textit{global links}.

Reducing the traffic forwarded over global links is advantageous for several reasons. First, global links are often longer than local links, resulting in higher communication latency~\cite{sc2020,frontier}. 
Second, because global links are typically oversubscribed, they can sustain only a limited number of concurrent communications, often causing congestion when multiple ranks from the same or different applications contend for the same link, leading to higher latency and lower throughput~\cite{gpcnet,sc2019,staci2018,gpugpuinterconnect}. This issue is amplified in oversubscribed low-diameter networks, where jobs are usually distributed across multiple groups. Such distribution helps arbitrary traffic patterns approximate uniform behavior, under which these topologies achieve high throughput~\cite{randompl1,randompl2,randompl3,slimfly,slurm_spread}, but it also increases global link utilization. Third, global links consume more energy than local links, making it crucial to reduce their usage~\cite{36462}.
Last, as the physical size of data centers and supercomputers approaches practical limits, tightly coupled applications run across multiple geographically distributed data centers~\cite{uno-tom,luo2024efficient,gemini_google,openai_yt,multidc}. In such scenarios, communication between distinct data centers or supercomputers must traverse even slower Wide-Area Network (WAN) links, and improving communication locality is critical. 

%In practice, we can imagine an HPC system as a collection of nodes grouped into local \textit{groups}, where communication within a group achieves significantly higher performance than communication across groups. These groups are often organized hierarchically. For example, a group could correspond to GPUs within a high-density compute node, where intra-node GPU communication achieves an order of magnitude or more higher performance than inter-node communication. Similarly, compute nodes may be clustered into fully connected groups (e.g., fully connected groups in a Dragonfly/Dragonfly+ network, or sub-trees in blocking fat trees). Even in this case, communication within a group (i.e., group) is typically faster than communication across groups, as the latter requires traversing global links that may be oversubscribed, leading to potential congestion (either by other ranks or other applications). This is even more relevant if we consider that the size of datacenters/supercomputers is reaching a limit, and that running tightly-coupled applications across multiple datacenters (connected through slower links) is starting to become a possibility~\cite{}. \dan{Mention energy etc}

For these reasons, running collectives on oversubscribed networks requires careful scheduling of communications to minimize traffic on global links and ensure scalability. Consider, for example, a \bcast operation implemented with a binomial tree. The root first sends the data to one process, and then each process forwards it to another, doubling the number of participants at each step until all processes have received the data. The algorithm remains correct as long as each process that has received the data sends it to one that has not, but different choices of destination processes can result in different amounts of traffic on global links.

%In general, when executed on $p$ processes (assuming, for simplicity, that $p$ is a power of two), the algorithm completes in $\log_2 p$ steps. The root, characterized by the highest communication load, transmits a total of $n \log_2 p$ bytes, where $n$ is the size of the broadcasted vector. 

%For the \bcast, they are traditionally built using either a \textit{distance-doubling} algorithm (as in Open MPI~\cite{ompi-bcast-binomial,ompi-binomial}) or a \textit{distance-halving} algorithm (as in MPICH~\cite{mpich-bcast-binomial}). 
%In both cases, the number of steps executed by the algorithm and the total number of bytes transferred remain the same. 
%and this has an impact on the number of bytes being transmitted on global links.
For instance, consider the scenario in Fig.~\ref{fig:bcast_motivation}, where eight nodes are connected through a 2:1 oversubscribed fat-tree. Each leaf switch has two downlinks and one uplink, serving two nodes per switch.
In a \textit{distance-doubling} binomial tree (as in Open MPI~\cite{ompi-bcast-binomial,ompi-binomial}, top part of Fig.~\ref{fig:bcast_motivation}), the distance between communicating ranks doubles at each step (e.g., rank 0 first sends its data to rank 1, then to rank 2). In contrast, in a \textit{distance-halving} tree (as in MPICH~\cite{mpich-bcast-binomial}, bottom part of the Fig.~\ref{fig:bcast_motivation}), the distance halves at each step (e.g., rank 0 first sends its data to rank 4, then to rank 2). 
In both cases, each rank that has already received the $n$ bytes of the vector forwards them to another rank. The distance-doubling \bcast forwards a total of $6n$ bytes over global links ($2n$ at step $1$ and $4n$ at step $2$), compared to only $3n$ bytes for the distance-halving variant.

\begin{figure}[htpb]
  \centering
  \includegraphics[width=\linewidth]{fig/motivation.pdf}
  \caption{Traffic on global links for a \bcast collective using \textit{distance-doubling} and \textit{distance-halving} binomial trees.}
  \Description{Traffic on global links for a \bcast collective using \textit{distance-doubling} and \textit{distance-halving} binomial trees.}
  \label{fig:bcast_motivation}
\end{figure}

\iffalse
\begin{figure*}[t]
  \centering
  \makebox[\textwidth][c]{
  \begin{subfigure}[b]{0.3\textwidth}
    \centering
    \includegraphics[height=3cm]{fig/motivation.pdf}
    \caption{First figure (wider)}
    \label{fig:fig1}
  \end{subfigure}
  \hspace{0.2\textwidth}
  \begin{subfigure}[b]{0.45\textwidth}
    \centering
    \includegraphics[height=3cm]{fig/trees_repr_a.pdf}
    \caption{Second figure (narrower)}
    \label{fig:fig2}
  \end{subfigure}
  }
  \caption{Two side-by-side figures with equal height}
  \label{fig:side-by-side}
\end{figure*}
\fi

This shows that, despite both algorithms requiring the same number of steps and transferring the same total data volume, the distance-halving variant is more efficient in reducing global link usage. 
%In general, this efficiency depends on how ranks are mapped to nodes, a topic we discuss further in Sec.~\ref{sec:bine}. 
Intuitively, closer communication is preferable when more ranks are involved in sending data.
Furthermore, if all links offer the same bandwidth, having two data transfers sharing a link can halve the effective injection bandwidth, as in step 1 of distance-doubling, between ranks 0-2 and 1-3. Though simplified for illustrative purposes, this example shows how the communication schedule affects inter-group traffic and performance. 
%We quantify this impact in practice on different supercomputers in Sec.~\ref{sec:evaluation}. 

Moreover, although ring algorithms reduce global link traffic compared to binomial trees and butterflies, their performance is suboptimal for large numbers of ranks and/or small to medium vector sizes~\cite{swing}.
Also, it is important to note that existing hierarchical and locality-aware collective algorithms may not always be effective. These algorithms typically require prior knowledge of the number of processes per group and often assume that jobs run with the same number of processes in each group~\cite{10.1145/3555819.3555825,10.1007/11846802_16,846009,5160896,10.1109/CCGrid.2011.42}. While such assumptions are reasonable for the number of processes per node, they may not hold when considering the partitioning of processes across subtrees or Dragonfly groups. In fact, the scheduler's process-to-node allocation is not known in advance and is unlikely to result in an even distribution of processes across groups. 

Based on these insights, this paper contributes as follows:

\begin{itemize}[left=0pt]
\item We introduce \bine (\textit{\underline{bi}nomial \underline{ne}gabinary}) trees\footnote{A \textit{bine} is a twisting vine, reminding the way in which we build \bine trees (Sec.~\ref{sec:bine:halving:visual}).} , a novel way of constructing binomial trees (and butterflies) collective algorithms, reducing the distance between communicating processes and, consequently, the volume of data transmitted across groups (Sec.~\ref{sec:bine}-\ref{sec:butterfly}). \Bine trees are topology-agnostic and generic as binomial trees and butterflies, but additionally optimize communication locality. They are orthogonal to hierarchical algorithms and can be used in both hierarchical and flat algorithms.
\item We design and implement new collective algorithms based on \bine trees and butterflies for \allgather, \allreduce, \reducescatter, \alltoall, \bcast, \gat, \reduce, \scatter (Sec.~\ref{sec:collectives})\footnote{Reference implementations are available at \url{https://github.com/HLC-Lab/bine-trees}.}.
\item We evaluate \bine collectives on up to \num{8192} nodes (Sec.~\ref{sec:evaluation}), on four supercomputers with diverse network topologies: \lumi (Dragonfly), \leonardo (Dragonfly+), \mn (2:1 oversubscribed fat-tree), and \fugaku (torus). We include three different MPI implementations: Fujitsu MPI, Open MPI, and Cray MPICH, as well as NCCL. \bine trees consistently outperform state-of-the-art algorithms, improving performance by up to 80\% on \lumi and \mn, 50\% on \leonardo, and up to $5\times$ on \fugaku, while also reducing global link load by up to 33\%. 
These results demonstrate the effectiveness of \bine trees in optimizing collective communication, delivering substantial performance and scalability benefits across diverse networks and software stacks.
\end{itemize}
%\section{Motivation}
\ifshowextra
An extended version of this paper provides further details, proofs, analysis, and discussion~\cite{bine_arxiv}.
\fi

\section{Distance-Halving \Bine Trees}\label{sec:bine}
We now recall the construction of binomial trees (Sec.~\ref{sec:bine:binomial}), then introduce distance-halving \Bine trees visually (Sec.~\ref{sec:bine:halving:visual}) and formally (Sec.~\ref{sec:bine:halving:formal}), and discuss their advantages over binomial trees (Sec.~\ref{sec:bine:halving:advantages}).

%\Bine butterflies and distance-doubling trees will instead be presented in Sec.~\ref{sec:butterfly}.

\subsection{Binomial Trees Construction}\label{sec:bine:binomial}
As shown in Fig.~\ref{fig:binomial}, a distance-halving binomial tree of order 0 consists of a single node. A tree of order $k$ is built recursively by linking two trees of order $k-1$, making one the child of the other. When used in collective operations, each node maps to a rank, and each edge represents communication between ranks. Any mapping of ranks to nodes yields a valid algorithm, also for collectives involving computation (e.g., \reduce), as long as the operator is associative, which MPI assumes by default.

\begin{figure}[htpb]
  \centering
  \includegraphics[width=.935\linewidth]{fig/trees_repr.pdf}
  \caption{Distance-halving binomial tree construction.}
  \Description{Distance-halving binomial tree construction.}
  \label{fig:binomial}
\end{figure}

Although the view of binomial trees shown in the top part of Fig.~\ref{fig:binomial} highlights the recursive construction of the tree, it does not clearly illustrate how communication progresses step by step.
To make this explicit, we use the representation in the bottom part of Fig.~\ref{fig:binomial}, where ranks are arranged left to right and each row (indicated with a dashed horizontal line) corresponds to a communication step. Boxes indicate that the value is present in a rank’s memory. Vertical lines show a rank lifetime, while diagonal arrows represent message transfers between ranks.
For example, in a \bcast across four ranks (i.e., a tree of order 2) with rank 0 as root, rank 0 first sends the data to rank 2 (Fig.~\ref{fig:binomial}, \raisebox{-0.2em}{\includegraphics[height=1em]{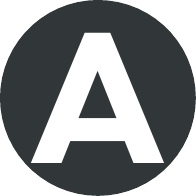}}). In the next step, rank 0 sends the data to rank 1 (Fig.~\ref{fig:binomial}, \raisebox{-0.2em}{\includegraphics[height=1em]{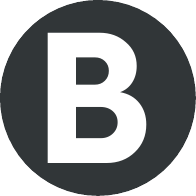}}), while rank 2 sends it to rank 3 (Fig.~\ref{fig:binomial}, \raisebox{-0.2em}{\includegraphics[height=1em]{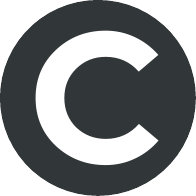}}).

\subsection{\Bine Trees Construction: Visual Intuition}\label{sec:bine:halving:visual}
%The key idea behind \Bine trees is to assign ranks to tree nodes to reduce the distance between communicating ranks (i.e., parent and child), thereby reducing inter-group traffic. For clarity and brevity, we assume the number of ranks $p$ is a power of two; when this is not the case, standard techniques used in binomial tree and butterfly algorithms apply (e.g., forwarding from excess ranks~\cite{ompi-allgather-recdoub,mpich-bcast-binomial}). We also assume for simplicity that the root of the tree is rank $0$. If the root, $t$, is different from $0$, we subtract $t$ from all rank identifiers modulo $p$ (i.e., rank $t$ behaves as if it was rank $0$).
The key idea of \Bine trees is to arrange ranks in a binomial tree so that communicating ranks are closer than in standard binomial trees. Reducing this distance increases the likelihood that communication remains within the same group, thereby benefiting from higher bandwidth and lower latency. 
To keep the algorithm simple and generic, and not tied to any specific topology, standard binomial trees do not use the actual physical distance between nodes (e.g., obtained from their location~\cite{6061055}); instead, they determine distances based on the difference in rank identifiers.

For example, in distance-halving binomial trees, rank $6$ first communicates with rank $4$ (distance $2$) and then with rank $7$ (distance $1$). 
This scheme reasonably approximates physical distance when ranks are mapped linearly across the system, which is often the case (e.g., under Slurm’s default \textit{block} distribution~\cite{slurm}). If this is not the case, a remapping to a \textit{block} distribution could be applied when the communicator is created, and the remapped rank identifiers used instead of the original ones. On the systems used in this work, for instance, hostnames are numbered consecutively across groups or subtrees, so it is sufficient to sort ranks by hostname.

Since our goal is to design an algorithm as simple and generic as standard binomial tree algorithms, we adopt a similar approach but use the \textit{modulo distance} between rank identifiers. Namely, we consider the $p$ ranks to be arranged in a circle $0,1,\ldots,p-1$ and define the distance between two ranks $r$ and $q$ as the minimum distance along the circle, i.e., $d(r, q) = \min((r - q) \bmod p,\,(q - r) \bmod p)$. 
To motivate this choice, consider three ranks ($0,1,2$), each in a different group. Standard binomial trees would regard the distance between ranks $0$ and $2$ as larger than that between ranks $0$ and $1$, whereas in practice, both involve inter-group communication and should therefore be treated as equal. Modular distance achieves this. Of course, counterexamples exist where the standard distance is more accurate. 
%For instance, with four ranks, $0$ and $1$ in one group and $2$ and $3$ in another, modular distance considers rank $0$ equally distant from ranks $1$ and $3$, although rank $1$ is in fact closer since it belongs to the same group. 
However, as shown in Secs.~\ref{sec:bine:halving:advantage:practice} and~\ref{sec:evaluation}, for jobs spanning multiple groups, as is often the case, modular distance provides a better approximation of the actual distance, enabling \Bine trees to outperform standard binomial trees in most scenarios.

%In other words, we assume that ranks are linearly mapped on a ring. In practice, if we consider two ranks under the same node to be at distance 0, two nodes under the same switch at distance 1, and so on, the difference $\bmod p$ between rank identifiers will always be greater than or equal to the actual distance.

%Instead, we assume that the difference between rank identifiers reflects their physical distance. That is, given any three ranks $i < j < k$, we can never have $i$ and $k$ in a group, and $j$ in a different one. For example, if ranks $1$ and $4$ are in the same group, then also ranks $2$ and $3$ must be in that group. This is always the case when ranks are mapped linearly across the system (e.g., using Slurm's default \textit{block} distribution~\cite{slurm}). If this is not the case, a remapping to a \textit{block} distribution can be computed when the communicator is created, and the remapped rank identifiers can be used in place of the actual identifiers. For example, on the systems considered in this work, hostnames are numbered consecutively across groups/subtrees, and it is enough to sort ranks according to their hostname. Importantly, we are not assuming any specific placement of ranks; we only assume that it is possible to virtually reorder them in a way that satisfies the condition above. 
%The concept of \textit{``distance''} heavily depends on the system's characteristics. For example, we might consider two nodes in the same group to be closer than two nodes in different groups. 

\begin{figure}[htpb]
    \centering
    \includegraphics[width=\columnwidth]{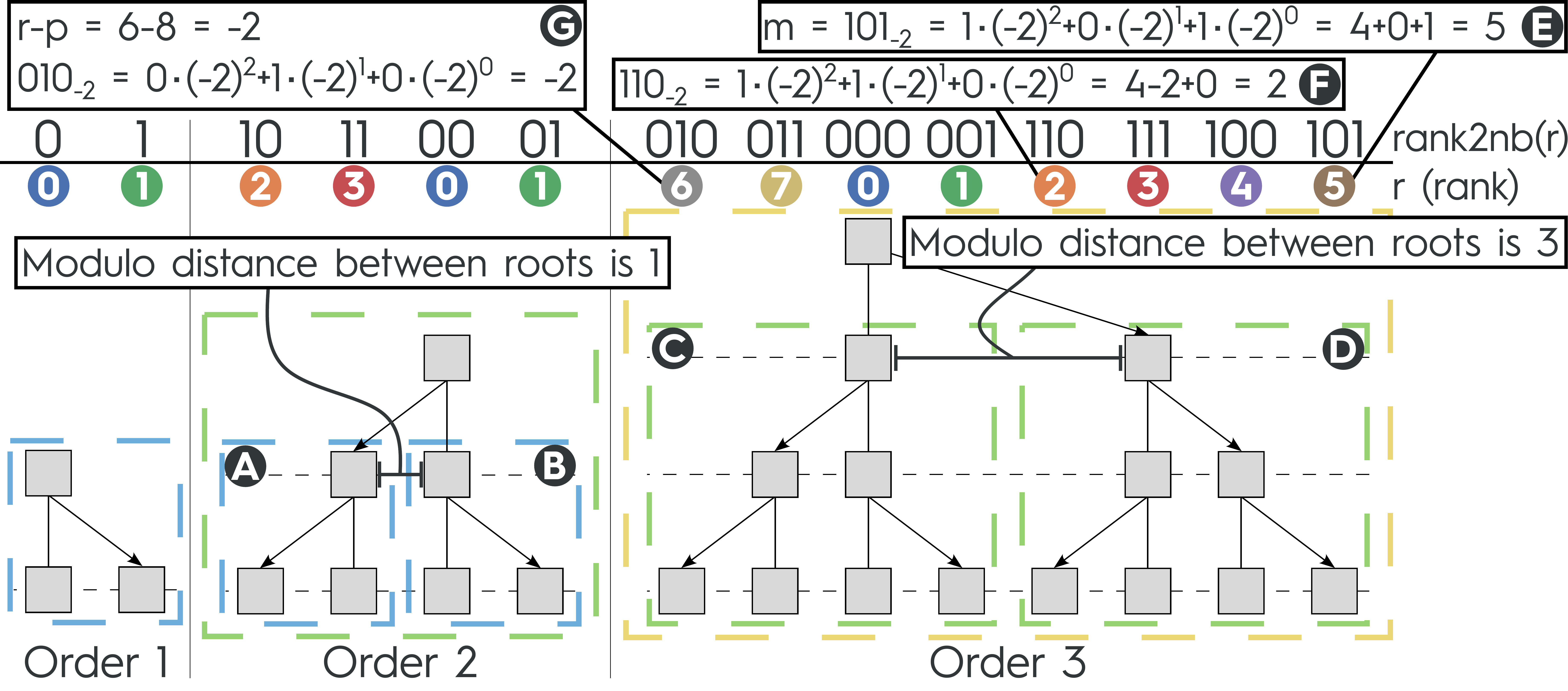}
    \caption{Distance-halving \bine trees construction.}
    \Description{Distance-halving \bine trees construction.}
    \label{fig:reduce_construction:small}
\end{figure}

We construct a distance-halving \Bine tree recursively. A two-node \Bine tree (Fig.~\ref{fig:reduce_construction:small}, order 1) is identical to a standard binomial tree. To build a four-node (i.e., order 2) \Bine tree, we connect an order 1 \Bine tree (Fig.~\ref{fig:reduce_construction:small}, \raisebox{-0.2em}{\includegraphics[height=1em]{fig/marker_b.pdf}}), to a mirrored order 1 \Bine tree (i.e., reflected on the vertical axis) and placed to its left (Fig.~\ref{fig:reduce_construction:small}, \raisebox{-0.2em}{\includegraphics[height=1em]{fig/marker_a.pdf}}). The mirrored tree is placed to the left rather than to the right to keep a shorter modulo distance between the roots of the subtrees (one rather than two as in a binomial tree -- see Fig.~\ref{fig:binomial}, \raisebox{-0.2em}{\includegraphics[height=1em]{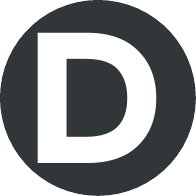}}). Ranks are numbered left to right starting from the root, wrapping around once the rightmost node of the tree is reached (e.g., on the 4-node tree, ranks $2$ and $3$ are those on the left of the root).

The process generalizes recursively: an eight-node \Bine tree is constructed by combining an order-2 \Bine tree (Fig.~\ref{fig:reduce_construction:small}, \raisebox{-0.2em}{\includegraphics[height=1em]{fig/marker_c.pdf}}) and a mirrored order-2 \Bine tree (Fig.~\ref{fig:reduce_construction:small}, \raisebox{-0.2em}{\includegraphics[height=1em]{fig/marker_d.pdf}}), placed side by side and arranged so that the modulo distance between the roots is as small as possible. In this case, placing the new four-node subtree on the right (rather than on the left) yields a root-to-root modulo distance of three, compared to five in the alternative placement. For comparison, the corresponding binomial tree would result in a modulo distance of four (Fig.~\ref{fig:binomial}, \raisebox{-0.2em}{\includegraphics[height=1em]{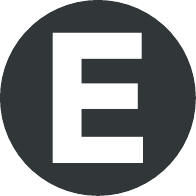}}). Appendix~\ref{apdx:nonp2} discusses how to handle cases where the number of nodes is not a power of $2$. Moreover, if the root is rank $t \neq 0$, we apply a logical rotation by subtracting $t$ from all rank identifiers modulo $p$. 
%In other words, rank $t$ behaves as if it was rank $0$.

%As mentioned in Sec.~\ref{}, our goal is not to make any assumption on the specific topology or on how the ranks are partitioned across groups but rather to design an algorithm that can reduce such inter-groups traffic in as many cases as possible.

\subsection{\Bine Trees Construction: Formal Definition}\label{sec:bine:halving:formal}
We now provide a formal construction process for distance-halving \Bine trees. Table~\ref{tab:notation} summarizes the notation we use.

\begin{table}[htpb]
\footnotesize
  \caption{Notation used.}
  \label{tab:notation}  
\begin{tabular}{p{1.2cm}p{6.5cm}}
\toprule
\textbf{Symbol} & \textbf{Definition} \\ 
\midrule
$p$         & Number of ranks. \\ 
%$p'$        & Biggest power of two smaller than $p$. \\
$r$         & Rank identifier. \\ 
$s$         & Number of steps in a binomial or \bine tree collective ($s = \log_2 p$). \\
%$m$         & Largest positive negabinary number on $s$ bits. \\
$rank2nb(r)$ & Convert rank identifier $r$ to its negabinary representation. \\ 
$nb2rank(r)$ & Get rank identifier $r$ from its negabinary representation. \\ 
\bottomrule
\end{tabular}
%\vspace{-2em}
\end{table}

\subsubsection{\textbf{Ranks Representation}}

In \bine trees, each rank is assigned a \textit{negabinary} (i.e., negative-base) representation, where numbers are expressed as sums of powers of $-2$ instead of $2$, as in standard binary. For example, the number $2$ is represented as $110_{-2}$, since $1 \cdot (-2)^2 + 1 \cdot (-2)^1 + 0 \cdot (-2)^0 = 4 - 2 = 2$. Unlike standard binary, negabinary representations can encode both positive and negative integers (e.g., $011_{-2} = -1$). As a result, there is no one-to-one correspondence between non-negative ranks and their negabinary representation.
The largest positive number $m$ that can be represented with a fixed number of negabinary bits is obtained by setting ones in all even positions and zeros in all odd positions. This ensures that only even powers of $-2$, which contribute positive values, are included. For instance, on six bits $m = 010101_{-2} =16+4+1 = 21$.

The construction of \bine trees ensures that the rightmost rank is $m$. For example, for a 8-nodes tree we need three bits, and $m=101_{-2}=5$ (Fig.~\ref{fig:reduce_construction:small}, \raisebox{-0.2em}{\includegraphics[height=1em]{fig/marker_e.pdf}}). Therefore, all ranks in the range $[0, m]$ (to the right of rank $0$) can be represented using their positive negabinary value (e.g., Fig.~\ref{fig:reduce_construction:small}, \raisebox{-0.2em}{\includegraphics[height=1em]{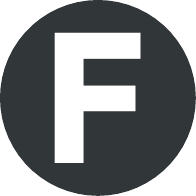}}). Ranks greater than $m$ (to the left of rank $0$) are represented using the negabinary representation of $r-p$. For instance, rank $r=6$ in an 8-node tree is represented as $6 - 8 = -2 = 010_{-2}$ (Fig.~\ref{fig:reduce_construction:small}, \raisebox{-0.2em}{\includegraphics[height=1em]{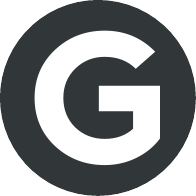}}).

We define the functions ${rank2nb}(r, p)$ and ${nb2rank}(r, p)$ to convert a rank identifier to its negabinary representation, and vice versa, for a collective on $p$ ranks. For example, ${rank2nb}(2, 8) = 110_{-2}$ and ${rank2nb}(6, 8) = 010_{-2}$. Both functions can be efficiently implemented using bit masking and an addition or subtraction. When the context is clear, we will omit the $p$ parameter and the $_{-2}$ base.

\subsubsection{\textbf{Determining the Communication Partner}}
\begin{figure}[htpb]
    \centering
    \includegraphics[width=\columnwidth]{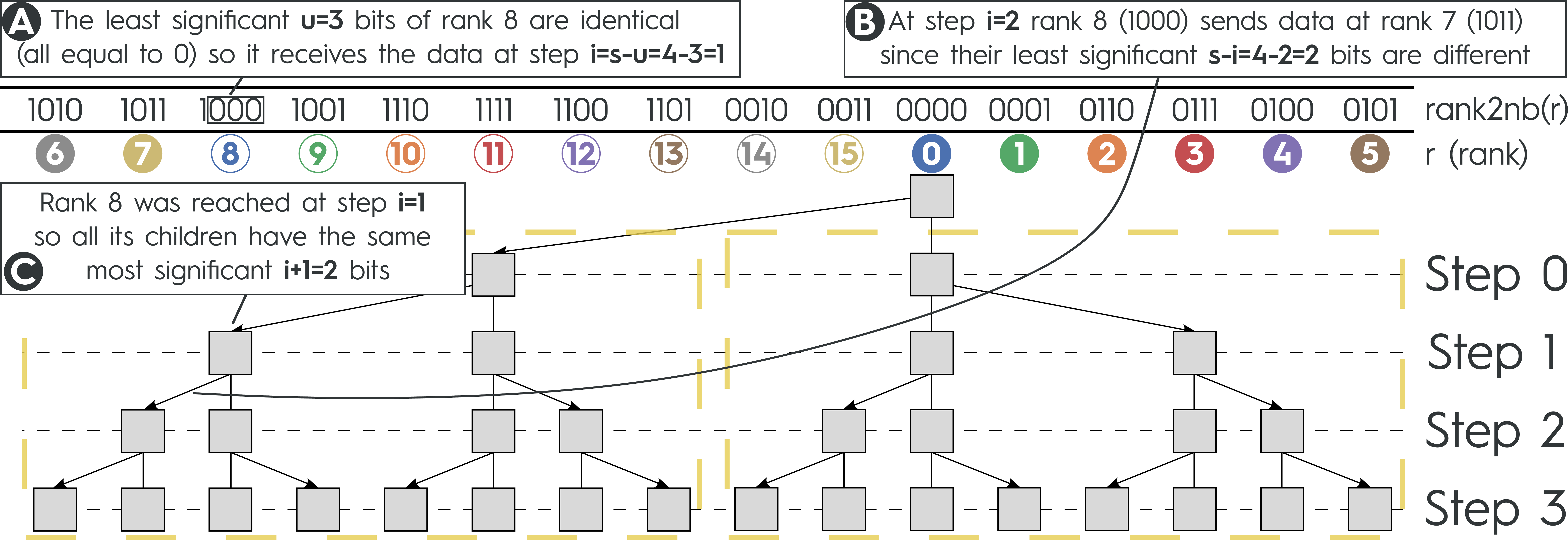}
    \caption{A 16-node (order 4) distance-halving \bine tree.}
    \Description{A 16-node (order 4) distance-halving \bine tree.}
    \label{fig:reduce_construction:large}
\end{figure}

In the \bcast, each rank must determine when to join the tree, or in other words, at which step it will receive data from its parent. We denote by $u$ the number of consecutive least significant bits that are equal to each other, starting from the least significant bit. E.g., for a 16-node \bine tree, $u = 3$ for $1000$, and $u=2$ for $1011$. In a distance-halving \bine tree \bcast, rank $r$ receives data at step $i = s - u$ (with $i \in [0, s - 1]$). 
As illustrated in Fig.~\ref{fig:reduce_construction:large} \raisebox{-0.2em}{\includegraphics[height=1em]{fig/marker_a.pdf}}, for rank $8$, we have $rank2nb(8) = 1000$, so it receives data at step $i = s - u = 4 - 3 = 1$.

%if the least significant $s-i$ bits of its negabinary representation are identical. Consider the 16-node \bine tree (order 4) shown in Fig.~\ref{fig:reduce_construction:large}, with each rank represented with $s = \log_2 16 = 4$ bits. For example, since $-4 \bmod 16 = 12$, we express $12$ as $rank2nb(-4) = 1100_{-2}$. Consequently, rank $12$ (i.e., $-4$) receives the data at step $2$, as its negabinary representation ($1100_{-2}$) has the least significant $s-i = 4-2 = 2$ bits all equal.

Once a rank receives the data, in each subsequent step $i$ it forwards the data to rank:
\begin{equation}\label{eq:qhalving}
q = {nb2rank}({rank2nb}(r) \oplus \overbrace{111\ldots1}^{s-i \:\text{bits}})
\end{equation}
with $\oplus$ denoting the XOR operation. In other words, at step $i$, each rank $r$ that has already received the data will send it to a rank $q$ whose negabinary representation differs in the least significant $s-i$ bits. For example, as shown in Fig.~\ref{fig:reduce_construction:large} \raisebox{-0.2em}{\includegraphics[height=1em]{fig/marker_b.pdf}}, at step $i=2$, rank $8$ sends data at rank $7$ since their negabinary representations ($1000$ and $1011$, respectively), differ for the least significant $s-i=4-2$ bits. For clarity, the complete MPI code of the distance-halving \Bine tree algorithm for the \bcast is provided in Listing~\ref{lst:bcast}. Notably, the implementation is no more complex than that of standard binomial trees~\cite{mpich-bcast-binomial,ompi-bcast-binomial}. Appendix~\ref{apdx:nonp2} discusses how to handle cases where the number of nodes is not a power of $2$.

\begin{listing}[htbp]
\includegraphics[width=\linewidth]{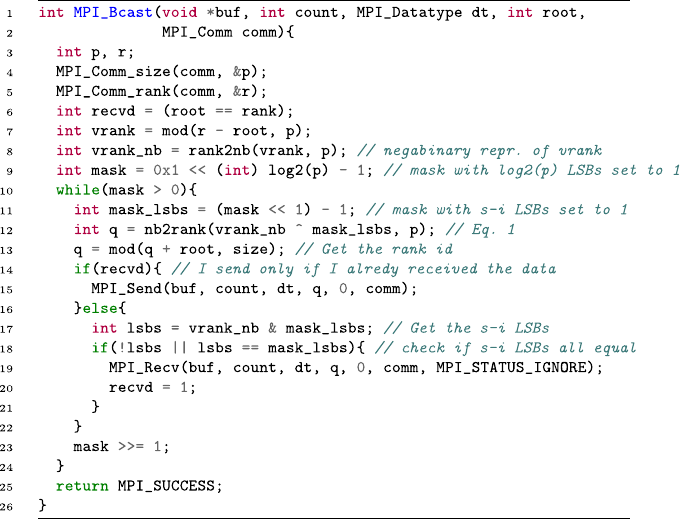}
    \caption{\Bine tree distance-halving \bcast code.}
    \label{lst:bcast}
\end{listing}

Correctness can be easily demonstrated.
%For example, rank $4$ receives the data via $0 \to 3 \to 4$, corresponding to the XOR combination $0000 \oplus 0111 \oplus 0011$. More generally, any $s$-bit number can be uniquely expressed as the XOR of a subset of $\{1, 11, 111, \ldots, \underbrace{111 \ldots 1}_{s \text{ bits}}\}$, which ensures correctness.
For example, rank $4$ receives the data via $0 \to 3 \to 4$, corresponding to the XOR combination $0000 \oplus 0111 \oplus 0011$. More generally, any $s$-bit number can be uniquely represented as the XOR of a subset of the binary/negabinary numbers $\{1, 11, 111, \ldots, \underbrace{111 \ldots 1}_{s \:\text{bits}}\}$, ensuring correctness.

\subsubsection{\textbf{Determining Ranks in a Subtree}}\label{sec:bine:halving:subtree}
As we will discuss in Sec.~\ref{sec:collectives}, collectives like \scatter (Sec.~\ref{sec:collectives:scatter}) require identifying all ranks in the subtree rooted at a given rank. After receiving data at step $i$, a rank applies XOR on its least significant bits to determine its next destination. This process recurses, and at each step, the $i+1$ most significant bits remain unchanged. As a result, all descendants of a rank share the same $i+1$ leading bits in their negabinary representation. For instance, in a 16-node \bine tree, rank $8$ is reached at step $i=1$, so all the ranks in the subtree rooted in $8$ share its two most significant bits (i.e., $10$, as shown in Fig.~\ref{fig:reduce_construction:large}, \raisebox{-0.2em}{\includegraphics[height=1em]{fig/marker_c.pdf}}).

\subsection{Advantages over Binomial Trees}\label{sec:bine:halving:advantages}
We know quantify the maximum reduction in global links traffic both theoretically (Sec.~\ref{sec:bine:halving:advantage:theory}) and empirically (Sec.~\ref{sec:bine:halving:advantage:practice}).

\subsubsection{\textbf{Theoretical Bound Computation}}\label{sec:bine:halving:advantage:theory}
To explain why we expect \bine trees to reduce the number of bytes on global links compared to standard binomial trees, we start by considering the distance between communicating ranks. In standard distance-halving binomial trees, two communicating ranks differ for the bit in position $s-i-1$ (i.e., the distance is $\delta_{{binomial}}(i) = 2^{s-i-1}$). In distance-halving \bine trees\footnote{Although we refer to it as a \textit{distance-halving} \bine tree for simplicity, the distances do not halve exactly; instead, they differ by at most $\pm1$ from the ideal halving.}, however, the negabinary representations of the two communicating ranks differ in the $s-i$ least significant bits, which are all equal to $1$ for one rank, and all equal to $0$ for the other rank. Thus, the distance between the two ranks at step $i$ is: $\delta_{bine}(i) = \lvert\sum_{j=0}^{s-i-1} (-2)^j\rvert = \lvert \frac{1}{3} - \frac{1}{3}(-2)^{s-i}\rvert \approx \frac{2^{s-i}}{3}$.

In binomial trees, the maximum distance between ranks is $p/2$, so the distance between communicating ranks coincides with their modular distance. This allows us to compare \Bine and standard binomial trees in terms of modular distance and to quantify the reduction in distance between communicating ranks using the ratio:
\begin{equation}\label{eq:bine_distance}
\frac{\delta_{bine}(i)}{\delta_{binomial}(i)} = \frac{2^{s-i}}{3\cdot2^{s-i-1}} = \frac{2}{3}
\end{equation}
This implies that in \bine trees, communicating ranks are at a $\sim33\%$ shorter modular distance compared to standard binomial trees, reducing the probability that communicating ranks are in different groups and communicate through global links. 

%\dan{actually we could quantify it numerically, just count how many connections go out of a group for a fixed group size, and then check when numinging < numrecdoub holds (i.e., for which group sizes)}

Quantifying the exact reduction in inter-group traffic is challenging, as it heavily depends on the specific topology and on how ranks are allocated to groups. However, we can estimate an upper bound by considering a worst-case scenario in which each group only has a single outgoing link to another group. 
Let us focus on a specific step $i$ and assume the group size equals $\delta_{binomial}(i)$, so that all communications in the standard binomial tree cross group boundaries. In this worst-case scenario, all $\delta_{binomial}(i)$ communications are directed to nodes in other groups, whereas for \bine trees this only happens for $\delta_{bine}(i)$ communications. Thus, the maximum reduction in global link traffic corresponds to the ratio $\delta_{bine}(i)/\delta_{binomial}(i)$, which, as shown in Eq.~\ref{eq:bine_distance}, is equal to $2/3$. In other words, \bine reduces by at most $33\%$ the traffic on global links.

\subsubsection{\textbf{Empirical Analysis}}\label{sec:bine:halving:advantage:practice}
We additionally validated our assumptions empirically by analyzing one-week job allocation data from the \leonardo~\cite{turisini2023leonardo} supercomputer and two-weeks data from the \lumi~\cite{lumisupercomputerLUMIsFull} supercomputer. Using Slurm commands (\texttt{squeue} and \texttt{scontrol}), we obtained the list of jobs running on the system and, for each job, the hostnames of the allocated nodes. We then identified the Dragonfly/Dragonfly+ group to which each node was assigned.
\ifshowextra
\else
On \leonardo, we relied on existing hostname-to-group mappings~\cite{leonardomap}, while on \lumi\ we extracted this information using the \texttt{sinfo -N -o "\%N \%f"} command.
\fi
With this data, we determined the group assignment of each rank for every job. Based on this mapping we estimated, for each job, the global traffic of an \allreduce operation using standard binomial and \bine trees, and computed the resulting traffic reduction achieved by \bine across allocations.

%For \leonardo, we used the available mapping data between hostnames and the Dragonfly+ groups~\cite{leonardomap} to calculate the number of bytes crossing global links during an \allreduce butterfly operation performed by both standard binomial trees and \bine trees (as discussed in Sec.~\ref{sec:bine:butterfly}, butterflies can be viewed as the overlay of multiple trees). On \lumi, we instead obtained the mapping of nodes to Dragonfly groups through the \texttt{sinfo -N -o "\%N \%f"} command. Using this data, we then computed the traffic reduction percentage achieved by \bine for each allocation.

%\begin{figure}[htpb]
%    \centering
%    \includegraphics[width=\linewidth]{plots/sinfo_summary_min16.pdf}
%    \caption{Reduction in global link traffic on 686 \leonardo's allocations.}
%    \Description{Reduction in global link traffic on 686 \leonardo's allocations.}
%    \label{fig:leonardo:sinfo}
%\end{figure}

\begin{figure}[htpb]
    \centering
    \includegraphics[width=\linewidth]{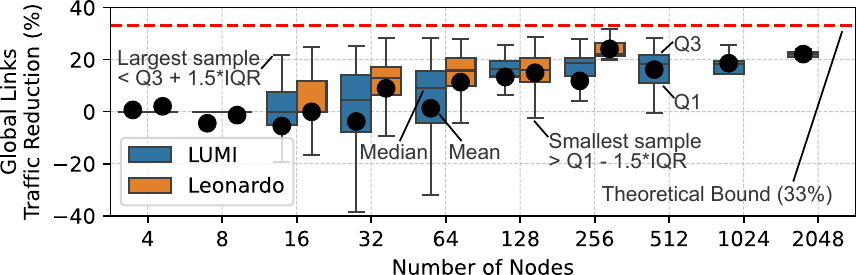}
    \caption{Reduction in global traffic on \leonardo and \lumi.}
    \Description{Reduction in global traffic on \leonardo and \lumi.}
    \label{fig:sinfo}
\end{figure}

We present this analysis in Fig.~\ref{fig:sinfo}, where we report the distribution of global traffic reduction across jobs using boxplots, grouped by the number of nodes. The data includes a total of \num{1116} jobs on \leonardo and \num{1914} jobs on \lumi. The largest observed jobs spanned up to \num{256} nodes on \leonardo and \num{2048} nodes on \lumi, matching the maximum job sizes allowed on the two systems.
%In each boxplot, the black dot indicates the mean, and the horizontal line within the box marks the median. The lower and upper edges of the box represent the first and third quartiles, respectively, while the whiskers extend to data points within 1.5 times the interquartile range (IQR). A dashed horizontal line at the top marks a $33\%$ reduction, which corresponds to the maximum possible reduction estimated through our model. 
For clarity, outliers are not shown in the figure.
%No outliers above the theoretical upper bound of $33\%$ (Eq.~\ref{eq:bine_distance}) were observed, while a few outliers reached values as low as $-200\%$, all of which occurred in small allocations (fewer than 64 nodes). As discussed in Sec.~\ref{sec:bine:halving:visual}, this happens because \bine minimizes the distance modulo $p$, and for a few small allocations it can increase the global traffic compared to binomial trees, although on average it is on par. In contrast, for larger allocations, \bine consistently outperforms standard binomial trees. Our findings show that the reduction in global link traffic achieved by \bine increases with the number of nodes. This trend is expected, as larger allocations tend to span more groups, leading to more opportunities for \bine to reduce inter-group communication.
No outliers above the theoretical upper bound of 33\% were observed, coherently with our model (Eq.~\ref{eq:bine_distance}). In a few cases, all occurring on less than 64 nodes, \bine trees increase global traffic. As discussed in Sec.\ref{sec:bine:halving:visual}, this happens because \bine minimizes distance modulo $p$ rather than the actual distance.
In contrast, for larger allocations, \bine consistently outperforms standard binomial trees. The reduction in global link traffic increases with the number of nodes, as larger jobs typically span more groups, creating more opportunities for \bine to reduce inter-group communication.

%we can empirically analyze the reduction in the number of inter-group transferred bytes by considering all the possible group sizes and locations. Namely, we consider all the possible starting positions of a group and all the possible group sizes, and we count, for each of those scenarios, the ratio between the bytes that exit the group when using standard binomial trees to those that exit the group when using \bine. We report the result of this analysis in Fig.~\ref{fig:reduce_analysis} for a reduce on 128 ranks. The analysis clearly shows that, although for some allocations \bine trees could send more inter-group bytes than traditional binomial trees, on average \bine trees send 30\% fewer bytes on inter-group links than binomial trees, which matches what estimated in Eq.~\ref{eq:bine_distance}. As we analyze in Sec.~\ref{sec:}, we run experiments on up to 2048 nodes on production supercomputers, observing similar reductions in the number of inter-group bytes, which eventually led to a 25\% performance improvement.

%\begin{figure}
%    \centering
%    \includegraphics[width=\linewidth]{fig/reduce_analysis.png}
%    \caption{Reduction in the number of bytes crossing different groups for different group configurations when using \bine trees. \dan{improve or remove it -- probably better to put here some real data already}}
%    \label{fig:reduce_analysis}
%\end{figure}

%\dan{replace with allocations observed on leonardo/lumi}

\section{\Bine Butterflies and Distance-Doubling Trees}\label{sec:butterfly}
\subsection{\Bine Butterflies}\label{sec:bine:butterfly}
%To explain how to build a butterfly using \bine trees, we consider as an example an \allreduce collective implemented through standard recursive doubling~\cite{}. In a recursive doubling collective, at each step each rank exchanges its buffer with another rank, and then aggregates the data received from the other rank. The communication pattern of a recursive doubling \allreduce is shown in Fig.~\ref{fig:}. In this case, at each step $i$ (starting from 0), each rank $r$ communicates with a rank $q = r \oplus 2^i$. This means that the distance between communicating ranks is equal to $2^i$. 

\begin{figure}[htpb]
  \centering
  \includegraphics[width=\linewidth]{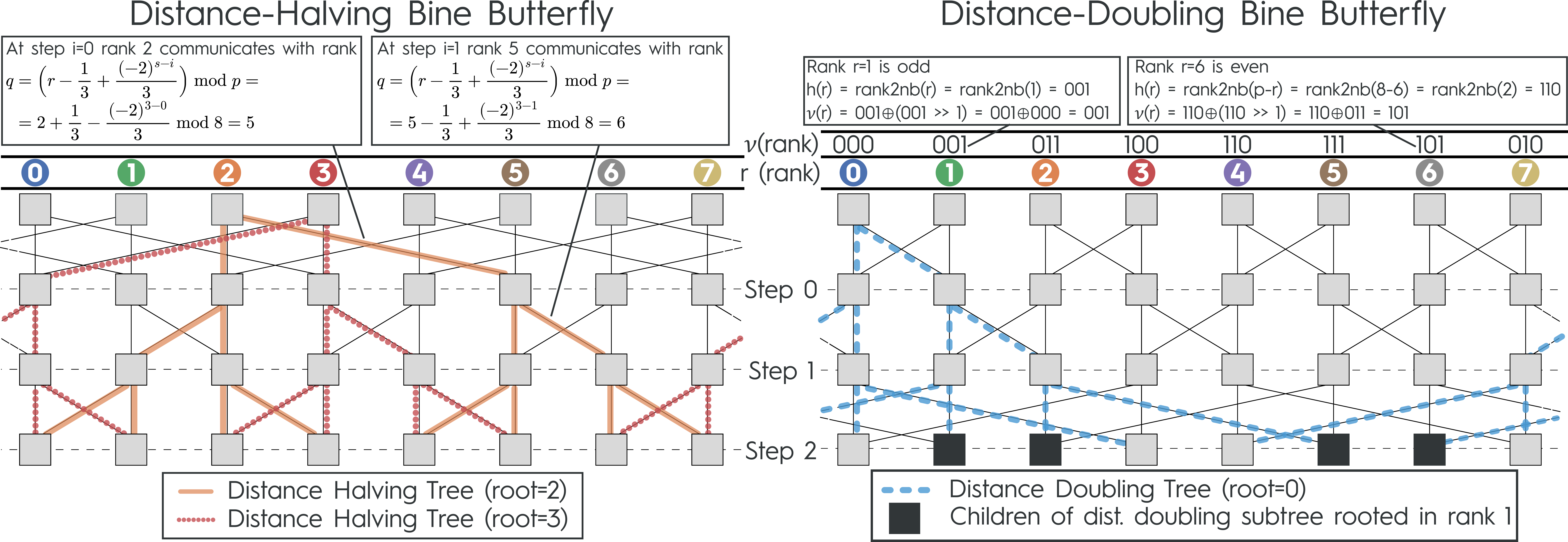}
  \caption{An 8-node distance-halving \bine butterfly (left), and distance-doubling \bine butterfly (right).}
    \Description{An 8-node distance-halving \bine butterfly (left), and distance-doubling \bine butterfly (right).}  
  \label{fig:butterfly}
\end{figure}

In a butterfly communication pattern, each rank exchanges data with another rank at every step. A \bine distance-halving butterfly can be seen as the superposition of multiple distance-halving \bine trees, each rooted at a different rank. The left part of Figure~\ref{fig:butterfly} illustrates this view, highlighting the trees rooted at ranks 2 and 3 with solid orange and dotted red lines, respectively. 

To guarantee correctness, these trees must be arranged so that the send and receive decisions are consistent across all ranks. We start from the distance-halving tree rooted in rank 0. As discussed in Sec.~\ref {sec:bine:halving:formal}, at step $i$ rank $0$ communicates with a rank whose negabinary representation has the least significant $s-i$ bits set to 1, and the remaining bits set to 0, i.e., with rank:
\begin{equation}
q = \Bigg(\sum_{j=0}^{s-i-1}(-2)^j\Bigg) \bmod p = \Big(\frac{1}{3}-\frac{(-2)^{s-i}}{3}\Big) \bmod p
\end{equation}
Even ranks $r$ communicate as rank $0$, but rotated to the right by $r$ positions. Moreover, since each rank communicates with another one whose negabinary representation differs in the last $i + 1$ bits, even ranks always communicate with odd ranks, and vice versa. Consequently, odd ranks $r$ communicate in a mirrored way with respect to the even rank $r-1$. Thus, the communication partner $q$ for a rank $r$ at step $i$ can be determined as:

\begin{equation}
q = \begin{cases}  
    \Big(r + \frac{1}{3}-\frac{(-2)^{s-i}}{3}\Big) \bmod p & \text{if } r \text{ is even} \\
    \Big(r - \frac{1}{3}+\frac{(-2)^{s-i}}{3}\Big) \bmod p & \text{if } r \text{ is odd}
\end{cases}
\end{equation}

%Therefore, if an even rank $r$ communicates with a rank $q = (r + \delta_{bine}(i)) \bmod p$ at step $i$, rank $r+1$ communicates with a rank $q = ((r+1) - \delta_{bine}(i)) \bmod p$.
%The tree rooted in rank $r$ is equal to the tree rooted in rank $0$, shifted ahead by $r$ positions if $r$ is even, or by $-r$ positions if $r$ is odd. In this case, we can see that the distance between communicating ranks is always equal to one. This is because the negabinary representation of the ranks involved in the communication differs for the least significant bit.

Since the distance between communicating ranks in \bine butterflies matches that of \bine trees, distance-halving \bine butterflies reduce global traffic by up to $33\%$ compared to standard distance-halving butterflies. Distance-halving butterflies are advantageous for collectives that increase the size of the exchanged data at each step, such as \allgather, since in the last steps, where more data is transmitted, communication happens between closer nodes.
Alternatively, it is possible to build a distance-doubling \bine butterfly by inverting the communication order, as shown in the right part of Fig.~\ref{fig:butterfly}. This is preferable for collectives that exchange more data in the first steps, such as \reducescatter (Sec.~\ref{sec:collectives:reducescatter_allgather}). Appendix~\ref{apdx:nonp2} discusses how to handle cases where the number of nodes is not a power of $2$.

\subsection{Distance-Doubling \Bine Trees}\label{sec:bine:doubling}
%We introduced the \bine butterfly as the overlay of multiple distance-halving \bine trees, each rooted in a different node, with roots at the top of the butterfly. However, we can also consider it as the overlay of multiple distance-doubling \bine trees, each rooted in a different node, with roots at the bottom of the butterfly. For example, we show with dashed blue lines in Fig.~\ref{fig:butterfly} the distance-doubling tree rooted in rank $0$. Differently from distance-halving \bine trees, in this case, we assign to odd ranks their negabinary representation $rank2nb(r)$, whereas we assign to even ranks the negabinary representation $rank2nb(-r)$. A rank joins rank 0's tree at step $i$, if its highest bit set to 1 is in position $i$ (e.g., rank 1 joins at step 0). For trees rooted in other ranks $t$, we adopt the same approach as in distance-halving trees, rotating rank identifiers by $t$ positions.

%We introduced the \bine butterfly as the overlay of multiple distance-halving \bine trees. However, we can also consider it as the overlay of multiple distance-doubling \bine trees, each rooted in a different node, as shown in the right part of Fig.~\ref{fig:butterfly}, where we highlight with dashed blue linesthe distance-doubling tree rooted in rank $0$.
%As with the distance-halving \bine tree, we first describe the case where the root is at rank 0. For a tree rooted at any other rank $t$, we follow the same approach as in distance-halving trees, by rotating rank identifiers to the right by $t$ positions.
We now describe the construction of distance-doubling \Bine trees, using as a reference the one rooted in $0$, shown with dashed blue lines in the right part of Fig.~\ref{fig:butterfly}.

\subsubsection{\textbf{Ranks Representation}}\label{sec:bine:doubling:representation}
We assign to each rank $r$ a value $h(r,p) = rank2nb(p-r, p)$ for $r$ even, and $h(r, p) = rank2nb(r, p)$ for $r$ odd. The only exception is $h(0, p) = 0$. Then, we assign to each rank $r$ a value $\nu(r,p) = h(r,p) \oplus (h(r,p) >> 1)$, with $>>$ being the logical right shift. When clear from the context, we omit $p$. 

\subsubsection{\textbf{Determining the Communication Partner}}\label{sec:bine:doubling:partner}
We describe the tree rooted at rank 0. For trees rooted at any other rank $t$, rank identifiers are rotated to the right by $t$ positions.
The communication partner of a rank $r$ at step $j$ is the rank $q$ such that $\nu(q) = \nu(r) \oplus 2^j$ (i.e., a rank that differs in the $j$-th bit in the $\nu$ representation). A rank communicates only after it receives the data from its parent. This happens at step $i$, where $i$ is the position of the highest set bit in $\nu(r)$. In the example in the right part of Fig.~\ref{fig:butterfly}, rank $2$ receives the data from its parent at step $1$, since the most significant bit set to $1$ in $\nu(2) = 011$ is in position $1$. Then, at step $2$, it sends the data to a rank whose $q$ such that $\nu(q) = 011 \oplus 2^2 = 011 \oplus 100 = 111$ (i.e., rank $5$). In other words, the algorithm operates as the standard binomial tree algorithm, but using $\nu(r)$ instead of $r$, thus reducing the modular distance by $33\%$
\ifshowextra
~\cite{bine_arxiv}.
\else
(see Appendix~\ref{apdx:dist-doubling-trees}). %.
\fi

\subsubsection{\textbf{Determining Ranks in a Subtree}}\label{sec:bine:doubling:subtree}
%Distance-doubling \bine trees require some attention for those collectives where each rank sends or receives at each step only parts of its vector (e.g., in a \scatter). Let us suppose to use the distance-doubling \bine tree highlighted in the right part Fig.~\ref{fig:butterfly} for a \scatter collective. Also, we consider the data to be divided in \quotes{\textit{blocks}} (one per rank). In a \scatter, each block $b$ must reach rank $b$ at the end of the collective. Thus, in our example, at step $0$, when rank $0$ sends part of its data to rank $1$, it needs to send not only rank $1$'s block, but also all the blocks of the nodes belonging to the subtree rooted in $1$ (i.e., $1$, $2$, $5$, and $6$). It is worth remarking that, differently from distance-halving trees, nodes in distance-doubling subtrees are not contiguous.
All ranks in a subtree rooted at $r$ have the same $i+1$ least significant bits in their $\nu$ representation, where $i$ is the step at which $r$ received the data from its parent. Indeed, after receiving the data at step $i$, rank $r$ applies XOR on the bits in positions greater than $i$. This process recurses, and at each step the $i+1$ least significant bits remain unchanged. 
E.g., rank $1$ receives data at step $i=0$, and its $i+1=1$ least significant bit in $\nu(1)$ is set. Consequently, all its descendants are ranks $q$ whose $\nu(q)$ also have the least significant bit set.
Nodes in distance-doubling \Bine subtrees are not contiguous, which may require transmitting non-contiguous data for some collectives (Sec.~\ref{sec:bine:noncontiguous}).

\section{Collectives Design}\label{sec:collectives}
%We first describe the algorithms assuming compute nodes are organized according to a flat structure, and we then generalize them for the case where they are arranged in a multi-dimensional torus. \dan{TODO: Explain better -- and avoid confusion between logical and physical topology}
%In this section we discuss how to use \bine trees and butterflies to implement \gat (Sec.\ref{sec:collectives:gather}), \scatter (Sec.\ref{sec:collectives:scatter}), \reducescatter and \allgather (Sec.\ref{sec:collectives:reducescatter_allgather}), \allreduce (Sec.\ref{sec:collectives:allreduce}), \alltoall (Sec.\ref{sec:collectives:alltoall}), and \bcast and \reduce (Sec.~\ref{sec:collectives:bcastreduce}).
%We would like to remark that all the algorithms we design are not hierarchical and do not make any assumption on the number of processes per node, the number of nodes per group, or the physical topology (except for the size of the dimensions in torus networks \dan{check}). Instead, they reduce the distance between communicating nodes so that, regardless of the allocation, they reduce the number of bytes going through global links. 
%As we discuss in Sec.~\ref{sec:} \dan{TODO}, these algorithms can be easily extended to consider hierarchical networks (e.g., for high-density nodes with multiple GPUs connected through NVLink~\cite{} or InfinityFabric~\cite{}). 

\subsection{Gather}\label{sec:collectives:gather}
\begin{figure}[htpb]
    \centering
    \includegraphics[width=\linewidth]{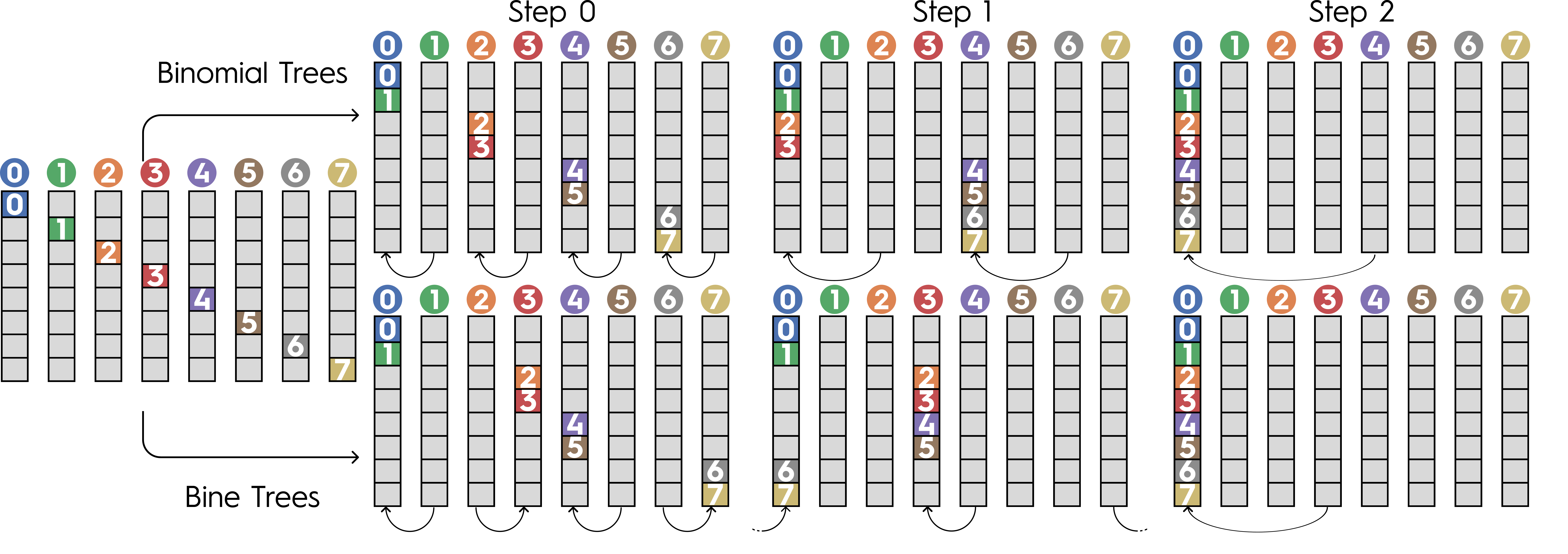}
    \caption{8-node \gat with binomial and \bine trees.}
    \Description{An 8-node \gat with standard binomial trees and \bine trees.}
    \label{fig:gather}
\end{figure}

As in standard binomial trees, each rank gathers data from another rank at each step, halving the number of active ranks until only the root remains. 
%After a rank sends its data, it will not participate in the subsequent communication steps. The amount of data sent and received by each rank doubles at each step, resulting in a constant total volume of data exchanged across steps. Consequently, the total number of inter-group bytes is unaffected by whether distance-halving or distance-doubling trees are used. 
We illustrate in Fig.~\ref{fig:gather}, a binomial and a \bine \gat over eight ranks, both using distance-halving trees. %We first assume the number of ranks is a power of two, and later generalize.
%In the first step, rank 0 receives data from rank 1, which is the only rank that has a negabinary representation differing from $00$ in the last two bits. In the second step, rank 0 receives data from rank 3, which is the only rank that has a negabinary representation differing from $00$ in the last three bits. In the third step, rank 0 receives data from rank 7, which is the only rank that has a negabinary representation differing from $00$ in the last four bits.
Data is divided into one \quotes{\textit{block}} per rank. Let $[a, b]$ denote the range of blocks a rank holds at any time. When sending, a rank transmits all the blocks it currently has. In a standard binomial \gat, if a rank holds $k$ blocks, it always receives the next $k$ blocks: that is, $[b+1, b+k]$. E.g., at step 1, rank 4 has blocks $[4, 5]$ and receives $[6, 7]$.

In \bine trees, however, the range may extend in either direction. A rank holding $[a, b]$ might receive either $[(b+1) \bmod p, (b+k) \bmod p]$ or $[(a-k) \bmod p, (a-1) \bmod p]$, extending its buffer upward or downward. For example, at step 1, rank 0 with blocks $[0, 1]$ receives $[6, 7]$ from rank 7, extending its buffer upward. In general, even ranks start extending downward, odd ranks upward, alternating direction at each step (e.g., rank 0 extends downward at step 0, upward at step 1, and downward again at step 2). 
%This might requires receiving two non-contiguous sub-buffers, and can be done either by sending two non-contiguous sub-blocks, using MPI datatypes, doing two separate sends, or manually copying the two parts of the buffer in a temporary buffer before transmission. We show in our evaluation in Sec.~\ref{} that this does not introduce any overhead compared to the standard binomial tree \gat, and that it is actually faster than the standard binomial tree \gat \dan{TODO:CHECK}. 

%\paragraph{\textbf{Generalization to an Arbitrary Number of Ranks}}
%If the number of ranks $p$ is not a power of two, we adopt a strategy similar to the standard distance-halving \gat. Let $p' = 2^{\lfloor\log_2p\rfloor}$ be the largest power of two smaller than $p$. The first step is executed only by the first $2(p - p')$ ranks, where each odd rank $r$ sends its data to rank $r-1$, effectively reducing the number of active ranks to $p'$. These $p'$ ranks then compute their negabinary representations (on $s = \log_2 p'$ bits) and proceed with the remaining $s$ steps as in the power-of-two case. When receiving data, they account for the fact that the first $p - p'$ positions of the buffer contain twice as many blocks, as they include data from the excluded $(p - p')$ ranks.

\subsection{Scatter}\label{sec:collectives:scatter}
For the \scatter we can use the opposite process of that used for the \gat. In this case, at each step, each rank sends either the bottom or the top half of its remaining buffer $[a, b]$. I.e., if a rank has $k$ blocks of data, it can either send the blocks $[a, (a+\frac{k}{2}) \bmod p]$, or the blocks $[(b - \frac{k}{2}) \bmod p, b]$ For example, since in the \gat rank 0 received in the last step the sub-buffer with range $[2, 5]$, then it means that in the \scatter it needs to send the sub-buffer with range $[2, 5]$ in the first step. Thus, we need to know what would have been the values of $[a, b]$ in the last step of the \gat. 

In the \gat, even ranks extended their buffer by adding $2^0 + 2^2 + 2^4 + \ldots$ to $b$ and subtracting $2^1 + 2^3 + 2^5 + \ldots$ from $a$. Odd ranks did the opposite: they subtracted $2^0 + 2^2 + 2^4 + \ldots$ from $a$ and added $2^1 + 2^3 + 2^5 + \ldots$ to $b$. This corresponds to adding or subtracting the binary numbers $0101\ldots0101_{2}$ or $1010\ldots1010_{2}$ to $a$ or $b$. Thus, it is enough to add and subtract those quantities to the rank identifier $r$ to find the starting $[a, b]$ for the \scatter, depending on whether the rank is even or odd. For example, in the first step of the \scatter, rank $0$ has $[a, b] = [6, 5]$. Then, at each step, either $b$ is decreased by $\frac{k}{2}$ or $a$ is increased by $\frac{k}{2}$, depending on whether the rank is even or odd. This process continues until $k = 1$.

%\paragraph{\textbf{Generalization to an Arbitrary Number of Ranks}}
%If the number of ranks $p$ is not a power of two, the \scatter can be handled similarly to the distance-doubling \bine tree \gat, but in reverse. %The data is initially scattered among $p'$ ranks, excluding the first $p-p'$ even ranks. Then, an additional final step involves the first $2(p - p')$ ranks, where each odd rank $r$ sends the data intended for rank $r-1$ to $r-1$.

\subsection{Reduce-Scatter and Allgather}\label{sec:collectives:reducescatter_allgather}
We discuss in this section the design of the \reducescatter collective. For the \allgather, it is enough to reverse the \reducescatter communication pattern. We consider a \reducescatter algorithm based on a vector-halving, distance-doubling butterfly pattern, using the \bine butterfly introduced in Sec.~\ref{sec:bine:butterfly}. 
At each step, each rank exchanges half of its current data with a peer, so the size of the exchanged data halves at each step. Since the communication volume is highest in the early steps, relying on a distance-doubling butterfly helps reduce global traffic. As in standard recursive halving \reducescatter, each rank sends a total of $n\frac{p-1}{p}$ bytes over $\log_2 p$ steps.

A distance-doubling butterfly \reducescatter can be conceptually viewed as the concurrent execution of a distance-doubling \scatter and a distance-halving \reduce. Focusing on the \scatter component, unlike the \scatter described in Sec.~\ref{sec:collectives:scatter}, a distance-doubling tree is used. While this reduces the volume of inter-group communication, it slightly complicates data selection: a rank $r$ must send to a rank $q$ all blocks associated with the ranks in the subtree rooted at $q$. As discussed in Sec.~\ref{sec:bine:doubling:subtree}, these subtrees consist of non-contiguous ranks, requiring the exchange of non-contiguous data blocks, which can significantly hinder performance, especially for small- to medium-sized vectors~\cite{sparbit}.

\subsubsection{\textbf{Dealing with Non-Contiguous Data}}\label{sec:bine:noncontiguous}
To mitigate the performance impact of transmitting non-contiguous data, we consider several possible options. One option is to use MPI datatypes to handle non-contiguous data or to explicitly copy the data to a temporary buffer before every transmission. However, this approach might still introduce a considerable overhead. Instead, for a \reducescatter, we consider the following alternatives, 
\ifshowextra
which we compare in the extended version of the paper~\cite{bine_arxiv}:
\else
which we compare in Appendix~\ref{apdx:noncontiguous}:
\fi

\begin{figure}[htpb]
    \centering
    \includegraphics[width=.65\linewidth]{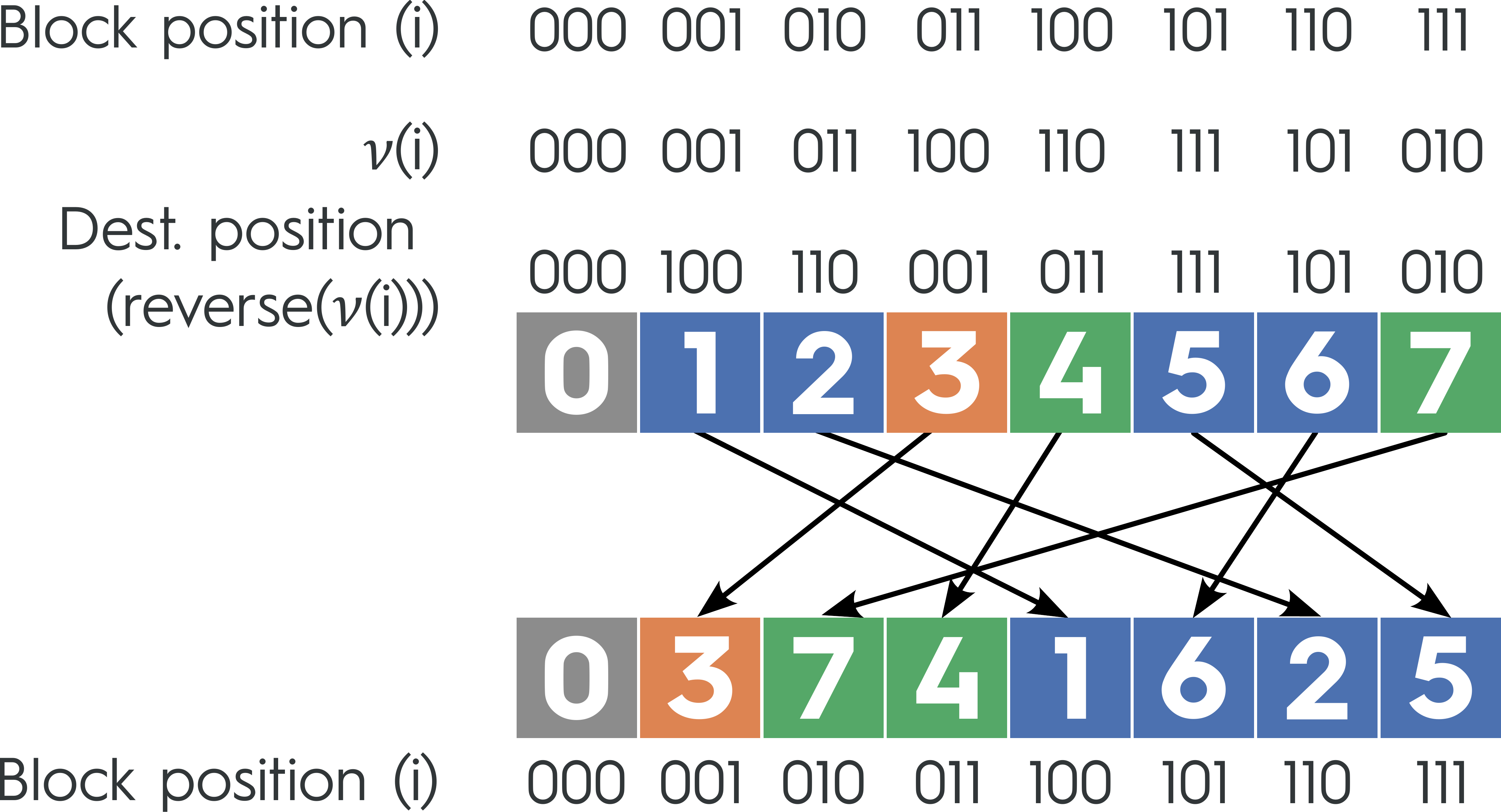}
    \caption{Permutation of non-contiguous data blocks to guarantee contiguous transmissions.}
    \Description{Permutation of non-contiguous data blocks to guarantee contiguous transmissions.}
    \label{fig:blocks_permute}
\end{figure}

\begin{description}[leftmargin=5pt, labelindent=0pt]
    \item[Block-by-block] Each block is transmitted independently. For small and medium-sized vectors, this approach increases the data transmission overhead. At the same time, it creates more opportunities for computation and communication overlap during reduction phases.
    \item[Permute] Data is first permuted so that subsequent transmissions operate on contiguous regions of memory. This optimization is only applicable when $p$ is a power of two. The permutation is obtained by moving each data block from position $i$ to position ${reverse}(\nu(i))$, where ${reverse}$ denotes the bit-reversal of $\nu(i)$. As discussed in Sec.~\ref{sec:bine:doubling:subtree}, all descendants of a given rank share a fixed number of least significant bits in their $\nu$ representation. By reversing $\nu(i)$, the transmitted blocks instead share a fixed number of most significant bits, making them contiguous in memory. For example, in Fig.~\ref{fig:blocks_permute}, at step~0 of the \reducescatter, rank~0 must send all blocks $i$ whose $\nu(i)$ representation has the least significant bit equal to~1 (see Sec.~\ref{sec:bine:doubling:subtree}). That is, it must send blocks $1$, $2$, $5$, and $6$. After applying the permutation that places each block $i$ at position ${reverse}(\nu(i))$, these four blocks occupy positions $4$–$7$, enabling contiguous transmission.
    \item[Send] Another option is to skip the permutation and let ranks transmit contiguous data as if the permutation had already been applied. In the example shown in Fig.~\ref{fig:blocks_permute}, this means that, in the first step, rank~0 would send the last four blocks of the vector (i.e., $4$, $5$, $6$, and $7$ instead of $1$, $6$, $2$, and $5$). As a result, at the end of the collective, each rank would hold the block belonging to another rank. For instance, after the \reducescatter, rank~1 would hold block~$4$ (since block $1$ was not moved in position $4$). 
    A final communication step can then be used to exchange and reorder the data correctly, with each rank~$r$ sending its block to rank $q = {reverse}(\nu(r))$. For example, rank~1 would send its block to rank $q = {reverse}(\nu(1)) = 4$.    
    It is worth noting that some collectives combine two operations (e.g., an \allreduce implemented as a \reducescatter followed by an \allgather). In such cases, even if ranks do not hold the correct blocks after the first phase, the subsequent collective (e.g., the \allgather) implicitly reverses the permutation and restores the correct order, thus avoiding the need for this additional communication step.        
    \item[Two Transmissions] A final option is to perform the \reducescatter using a distance-halving rather than a distance-doubling butterfly. While this increases traffic over globally oversubscribed links, it simplifies handling non-contiguous data. Specifically, similar to the distance-halving \scatter (Sec.~\ref{sec:collectives:scatter}), blocks are contiguous if the buffer is considered circular: any overflow at the end wraps around to the beginning. Thus, in practice this may require sending two segments: one at the end of the buffer and one at the beginning (for example, see rank~0's buffer at step~1 of the \bine tree in Fig.~\ref{fig:gather}).
\end{description}

For an \allgather, we have a symmetric scenario, i.e., in the \textit{permute} case, the permutation is done at the end, after all the ranks received the entire buffer, and in the \textit{send} case, the transmission to reorder the blocks is done before the actual steps of the collective operation.

\subsection{Allreduce and Alltoall}\label{sec:collectives:allreduce}
For the \allreduce, we use two algorithms: one for small and one for large vectors. As for standard binomial tree algorithms, the former performs fewer steps, whereas the latter performs more steps but transfers less data. For small vectors, we employ a standard recursive doubling \allreduce using \bine butterflies. For large vectors, we perform a \bine \reducescatter followed by a \bine \allgather. Both algorithms create a communication pattern similar to the Swing algorithm~\cite{swing}. However, unlike Swing, which always sends non-contiguous data~\cite{swing-git} and needs to issue multiple sends at each step, \Bine sends contiguous data (when $p$ is a power of two), leading to $2\times$ performance improvements (see Sec.~\ref{sec:evaluation:leonardo:sota}). %After the \reducescatter, ranks may have incorrect blocks, but the \allgather will reorder them. %Additionally, while Swing requires unrolling the entire tree beforehand to determine the data to send, the \bine tree allows on-the-fly decision making. We compare the \bine \allreduce with Swing in Sec.~\ref{}. \dan{does it sound incremental?}

%\paragraph{\textbf{Generalization to an Arbitrary Number of Ranks}}
%For the small vectors \allreduce we use a standard approach where the last $p-p'$ ranks send their data to the the first $p-p'$ ranks. Then, an allreduce is executed on $p'$ ranks and, eventually, the first $p-p'$ ranks send the fully aggregated data to the last $p-p'$ ranks. For large vectors, we can use the same approach used for \reducescatter and \allgather, thus not requiring to transmit any additional data, nor executing any additional step. 

Conceptually, the \alltoall can be viewed as a small vector \allreduce, where ranks send at each step $n/2$ bytes, and the received data is concatenated with their own rather than being aggregated. At each step, each rank moves the data it wants to keep to the left of its buffer and the data it needs to send to the right, similar to the rotations in Bruck's algorithm. Each rank tracks the block stored in each buffer position. At the end of the algorithm, a final permutation step is needed to put the data in the correct order.

\subsection{Reduce and Broadcast}\label{sec:collectives:bcastreduce}
%In the \reduce collective, one \textit{root} node aggregates a set of vectors (one per rank) using a standard or custom reduction operator. In the \bcast collective, data from a root node is sent to all the other nodes. 
For \bine \reduce and \bcast, we propose algorithms for both small and large vectors. The small vector algorithm uses distance-halving \bine trees for broadcasting data from the root to the leaves or reducing data from the leaves to the root. For large vectors, we implement the \reduce as a \reducescatter followed by a \gat, and the \bcast as a \scatter followed by an \allgather. The \reduce uses a distance-doubling \bine butterfly for the \reducescatter and a distance-halving \bine tree for the \gat. Data is transmitted contiguously, without the need for reordering, as the \gat inverts the block permutation done by the \reducescatter, placing the blocks back in the correct order. Similarly, the \bcast relies on distance-doubling \scatter and distance-halving \allgather.

%\paragraph{\textbf{Generalization to an Arbitrary Number of Ranks}}
%For small vectors, if the number of ranks $p$ is not a power of two, we use an approach similar to the standard binomial tree. We denote $p'$ as the largest power of two smaller than $p$ (i.e., $p' = 2^{\lfloor\log_2 p\rfloor}$). The first step is executed by the first $2(p-p')$ ranks, reducing the number of active ranks to $p'$. These ranks compute their negabinary representation and complete the remaining $\log_2 p'$ steps as described. For large vectors, the considerations for \reducescatter, \allgather, \scatter, and \gat apply.

%\dan{What about MVAPICH?}

%\dan{what about k-nomial trees?}

%\begin{figure}
%    \centering
%    \includegraphics[width=0.9\linewidth]{fig/binomial_tree_standard.png}
%    \caption{A binomial tree.}
%    \label{fig:binomial_tree_standard}
%\end{figure}

%\begin{figure}
%    \centering
%    \includegraphics[width=\linewidth]{fig/mapping}
%    \caption{Linear and round-robin mapping of ranks across nodes.}
%    \label{fig:mapping}
%\end{figure}

%\begin{figure}
%    \centering
%    \includegraphics[width=\linewidth]{fig/cartesian}
%    \caption{\name optimization for Cartesian topologies. \dan{find an easier/better way to explain. Also, green arrow should come from right rather than bottom.}}
%    \label{fig:cartesian}
%\end{figure}

%This, however, leads to transmitting non contiguous data blocks both in scatter and gather \dan{todo:elaborate more}

\section{Experimental Evaluation}\label{sec:evaluation}
We evaluate \bine trees on four supercomputers ranked among the top 11 in the Top500 list at the time of writing. Table~\ref{tab:systems} summarizes their network topologies and the MPI libraries used, as recommended by their respective supercomputing centers. Additional system-specific details appear in later sections. Our selection covers all four commonly used oversubscribed network topologies, providing a broad and representative evaluation. %Throughout the paper, we refer to both tree and butterfly variants simply as \Bine or binomial trees.

\begin{table}[htpb]
\footnotesize
  \caption{Systems used in the experimental evaluation. Top500 rankings taken from November 2024 list.}
  \label{tab:systems}  
\begin{tabular}{p{2.5cm}p{2.8cm}p{2cm}}
\toprule
\textbf{System (Top500 Rank)} & \textbf{Topology} & \textbf{MPI Library} \\
\midrule
\textbf{LUMI (\#8)} &  Dragonfly & Cray MPICH v8.1.29\\
\textbf{Leonardo (\#9)} &  Dragonfly+ & Open MPI v4.1.6 \\
\textbf{MareNostrum 5 (\#11)} & 2:1 Oversubscribed Fat Tree & Open MPI v4.1.5 \\
\textbf{Fugaku (\#6)} & 6D Torus & Fujitsu MPI v4.0.1 \\
\bottomrule
\end{tabular}
%\vspace{-1em}
\end{table}

\paragraph{\textbf{Implementation}}
We implemented all \Bine algorithms from Sec.~\ref{sec:collectives} on top of MPI, without modifying any MPI library. This choice was motivated by both portability (our target systems rely on four different MPI libraries, see Table~\ref{tab:systems}) and fairness, allowing for direct comparison against state-of-the-art algorithms using the same MPI backend. To validate this design choice, we implemented \Bine versions of \allreduce, \allgather, and \reducescatter within Open MPI and compared them to their counterparts built on top of MPI. The observed performance difference was below 3\%. Therefore, in the following, we report results using the \Bine implementations built on top of MPI. This was not the case on \fugaku, where, to concurrently use all the available NICs, our implementation uses the low-level uTofu library~\cite{utofu} (see Sec.~\ref{sec:evaluation:fugaku}). We implemented all \Bine algorithms in the PICO benchmarking framework~\cite{pico,pico-github}, which streamlines comparison with the collective algorithms provided by default in Open MPI and MPICH. Additionally, we provide self-contained reference implementations~\cite{bine-github}.
%Additionally, we ported Open MPI’s default binomial implementations to run on top of MPI for a fair baseline. In both cases,

\paragraph{\textbf{Algorithms Used in the Comparison}}
We compare \Bine trees against all the collective algorithms available in the MPI implementations on each system. To ensure a fair comparison, we evaluate \Bine both against the default algorithm selected by MPI and each available algorithm individually, by manually enforcing the selection. We also added other state-of-the-art algorithms from other MPI libraries or versions, such as the \textit{sparbit} \allgather algorithm~\cite{sparbit,ompi-allgather-sparbit} (not available on Open MPI v4.1.x), and the ring-based \allreduce algorithm (not available in MPICH 3.4a2, on which Cray MPICH v8.1.29 is based~\cite{mpichnoring}). Moreover, we implemented the Swing algorithm~\cite{swing} for \allgather, \reducescatter, and \allreduce, and the torus-optimized Bucket algorithm~\cite{bucket1,bucket2} (essentially a multi-dimensional ring) for the same collectives. For each system, we begin with a focused comparison of \bine and binomial trees, and then extend the evaluation to include all available algorithms.

\paragraph{\textbf{Benchmarking Methodology}}
For each collective, we measured performance across different node counts (without requesting any specific node placement) and vector sizes ranging from 32 B to 512 MiB. Due to space constraints, we report results only for power-of-two node counts. Results for non-power-of-two cases are consistent with those shown and are available in the reproducibility package.
All collectives operate on vectors of 32-bit integers. For each combination of node count and vector size, we executed each collective up to \num{20000} times for small vectors and five times for 512 MiB vectors. Most experiments were conducted with one process per node; performance with multiple processes per node is discussed in Sec.\ref{sec:discussion}. All vectors are located in host memory except for the experiments in Sec.~\ref{sec:discussion} where we analyze multi-GPU collectives. We followed standard benchmarking practices~\cite{htorbenchmarking}, by discarding the first 20\% of iterations as warm-up, and reporting for each iteration the maximum runtime across all ranks.

% We have Bine, SoTA, Binomial. We can have the following cases:
% - (x2) Sota > all: We just show sota name.
% - (x1) Bine > SoTA > Binomial: We show gain in perf and traffic -- We should also make it clear that this is a case where we actually improved binomial by a lot. It was losing and now it wins
% - (x1) Bine > Binomial > SoTA: Just show gain in perf and traffic -- binomial was already winning before
% - (x1) Binomial > Bine > SoTA: Show name, but we should also show by how much (usually is a few perc.). These cases are very rare I guess
% - (x1) Binomoial > SoTA > Bine: as before I guess -- to check

\subsection{Evaluation on LUMI}\label{sec:evaluation:lumi}
LUMI~\cite{lumisupercomputerLUMIsFull} relies on a 24-group Slingshot 11 Dragonfly network~\cite{sc2020}, with 124 nodes per group. We used the LUMI-G partition, where each node has a 64-core AMD EPYC\texttrademark{} Trento 7A53 CPU (organized into four NUMA domains with 128GB of DDR4 memory each), four 200Gb/s Cassini-1 NICs, and four AMD MI250x GPUs.

\subsubsection{\textbf{Comparison with Binomial Trees}}\label{sec:evaluation:lumi:bvb}
We compare \Bine and binomial trees in terms of both performance and the reduction in the number of bytes transmitted over global links. To assess this, we calculated the total bytes exchanged between ranks. By using the node locations from the \texttt{/etc/cray/xname} file, we identified the group each node belongs to and quantified the traffic forwarded between ranks assigned to different groups. %Due to space constraints, for \allreduce, \bcast, and \reduce we only report results for large-vector algorithms.

On LUMI, we conducted experiments using between \num{16} and \num{1024} nodes (the maximum allowed) spanning from one to 21 groups. As discussed in Sec.~\ref{sec:collectives:reducescatter_allgather}, \Bine’s \reducescatter and \allreduce algorithms may send multiple non-contiguous data blocks, rather than a single bulk transfer. This design improves the overlap between communication and computation in \reducescatter. In contrast, the standard MPICH binomial tree and butterfly implementations transmit large contiguous blocks, without exploiting such overlap. To ensure a fair comparison focused on algorithmic structure rather than implementation-level optimizations, we excluded these \Bine variants from the analysis. For the \alltoall collective, we compared \Bine with Bruck’s algorithm, as it is the closest to binomial algorithms, performing a number of steps logarithmic in $p$.

Given the large volume of data collected, across eight collectives, up to ten algorithms per collective, nine vector sizes, and up to seven node counts, for a total of over \num{5000} configurations, we present a summarized analysis in Table~\ref{tab:results:summary:lumi:bvb}\footnote{The complete set of detailed results is available in the reproducibility package.}. For each collective, we report the fraction of configurations (i.e., combinations of node count and vector size) in which \Bine outperforms binomial trees and vice versa (\quotes{\textit{\%Win/\%Loss}}). When \Bine is superior, we report both the average and maximum performance improvements over all the configurations. We compute the average using the geometric mean, which is more appropriate for summarizing ratios~\cite{htorbenchmarking}. Similarly, when binomial trees outperform \Bine, we report the corresponding average and maximum performance degradations. 

Lastly, we report both the average and maximum reduction in global-link traffic. Here, global links refer to those connecting switches in different Dragonfly groups. We assume packets traverse inter-group connections via minimal paths, an assumption that may not always hold in practice, especially in low-diameter networks such as Dragonfly~\cite{sc2020}. As a result, the reductions we report should be interpreted as lower bounds; the traffic savings induced by \Bine trees could be even greater than our estimation. %We consider\textit{sparbit}~\cite{sparbit,ompi-allgather-sparbit}.

\begin{table}[htpb]
\footnotesize
\centering
\caption{Comparison with Binomial Trees on LUMI}
\label{tab:results:summary:lumi:bvb}
\resizebox{.85\columnwidth}{!}{
\begin{tabular}{lccccc}
\toprule
\textbf{Coll.} & \textbf{\% Win} & \makecell{\textbf{Avg/Max}\\\textbf{Perf. Gain}} & \textbf{\% Loss} & \makecell{\textbf{Avg/Max}\\\textbf{Perf. Drop}} & \makecell{\textbf{Avg/Max}\\\textbf{Traffic Red.}}  \\
\midrule
Allreduce & 67\% & 7\%/67\% & 14\% & 9\%/13\% & 11\%/20\% \\
Allgather & 47\% & 19\%/177\% & 41\% & 11\%/17\% & 11\%/20\% \\
Red.-Scat. & 39\% & 21\%/143\% & 29\% & 13\%/29\% & 11\%/20\% \\
Alltoall & 94\% & 33\%/191\% & 2\% & 16\%/16\% & 15\%/20\% \\
Bcast & 67\% & 25\%/79\% & 27\% & 16\%/29\% & 88\%/94\% \\
Reduce & 87\% & 7\%/32\% & 5\% & 10\%/14\% & 10\%/19\% \\
Gather & 71\% & 13\%/47\% & 16\% & 13\%/32\% & 9\%/25\% \\
Scatter & 61\% & 7\%/30\% & 24\% & 9\%/20\% & 9\%/25\% \\
\bottomrule
\end{tabular}
}
\end{table}

Overall, \Bine outperforms standard binomial trees in more than 60\% of the configurations on most collectives. The only exceptions are \reducescatter and \allgather, where \Bine still performs better more often than it is outperformed by binomial trees. For \allgather, \Bine is superior in 47\% of the cases, while binomial trees outperform it in 41\%, with the remaining 12\% showing minimal differences (below 1\%). The smaller advantage for these two collectives is due to the additional data permutation step required by \Bine trees, either through local buffer shuffling or additional data communication (as discussed in Sec.~\ref{sec:collectives:reducescatter_allgather}). However, when \Bine outperforms binomial trees, the average performance gain is around 20\%, with a maximum improvement of up to 177\%.

Notably, for \alltoall, \Bine outperforms Bruck's algorithm in 94\% of the cases. Given that \alltoall typically involves large data transfers, reducing the traffic on global links becomes especially important. However, as we will discuss in Sec.~\ref{sec:evaluation:lumi:sota}, logarithmic \alltoall algorithms like Bruck's and \bine are typically only effective for small messages and large node count, while linear algorithms tend to perform better in the other cases. It is worth remarking that, as is typical for collective operations, the optimal algorithm depends on both the vector size and the number of nodes. As a result, no single algorithm consistently outperforms the others across all scenarios, and such trade-offs are therefore expected for all collectives.

\Bine reduces traffic on global links by an average of 10\% across all collectives, and up to 20\% at larger node counts, as more nodes lead to higher utilization of global links. The largest reduction, up to 94\%, is observed in \bcast. This is due to MPICH (and Open MPI, as shown later) using a large-vector \bcast based on a distance-halving \scatter, followed by a distance-doubling \allgather~\cite{mpich-bcast-sag,ompi-bcast-sag}. In contrast, \Bine trees reverse this process, significantly reducing traffic on global links, particularly in the \allgather phase.

\iffalse
\begin{figure*}[t]
  \centering
  \begin{subfigure}[t]{0.24\textwidth}
    \includegraphics[width=\linewidth]{plots/lumi_hm/allreduce/tasks_per_node_1_metric_mean_base_all_y_no_False.pdf}
    \caption{Allreduce}
    \label{fig:plot:lumi:allreduce}
  \end{subfigure}
  \begin{subfigure}[t]{0.24\textwidth}
    \includegraphics[width=\linewidth]{plots/lumi_hm/reduce_scatter/tasks_per_node_1_metric_mean_base_all_y_no_False.pdf}
    \caption{Reduce-Scatter}
    \label{fig:plot:lumi:reducescatter}
  \end{subfigure}
  \begin{subfigure}[t]{0.24\textwidth}
    \includegraphics[width=\linewidth]{plots/lumi_hm/reduce/tasks_per_node_1_metric_mean_base_all_y_no_False.pdf}
    \caption{Reduce}
    \label{fig:plot:lumi:reduce}
  \end{subfigure}
  \begin{subfigure}[t]{0.24\textwidth}
    \includegraphics[width=\linewidth]{plots/lumi_hm/gather/tasks_per_node_1_metric_mean_base_all_y_no_False.pdf}
    \caption{Gather}
    \label{fig:plot:lumi:gather}
  \end{subfigure}  
  \caption{Performance Analysis of \Bine Trees on LUMI. Each cell reports either a letter indicating the best-performing algorithm or, when \Bine is the best, the performance ratio between \Bine and the next-best state-of-the-art algorithm. N = binomial, P = pairwise, B = nonblocking.\dan{TODO: Percentage rather than 1.ax, exploit noylabel}}
  \label{fig:plot:lumi}
\end{figure*}
\fi

\subsubsection{\textbf{Comparison with Other State-of-the-Art Algorithms}}\label{sec:evaluation:lumi:sota}
To provide a complete comparison, we now compare \Bine with state-of-the-art algorithms (including binomial trees as well).

\paragraph{\textbf{Allreduce}} We present a detailed analysis of the \allreduce in Fig.~\ref{fig:plot:lumi:allreduce}, using a heatmap to display all combinations of node count and vector size. Each cell either indicates the best-performing algorithm with a single letter or, if \Bine is superior, shows the performance ratio between \Bine and the best state-of-the-art algorithm. \Bine outperforms other algorithms in almost all configurations, with peak performance gains of up to $1.62\times$ on medium-sized vectors and high node counts. %As node count increases, the ring algorithms experience performance degradation since due to their latency scaling linearly with node count, while traditional binomial trees and butterfly algorithms also start to suffer despite their logarithmic latency.
%This occurs because each rank in ring-based approaches sends data to more distant nodes, leading to higher communication across global links and resulting in congestion and latency penalties. In contrast, \Bine trees and butterflies maintain the logarithmic latency scaling of binomial and butterfly algorithms while reducing the amount of data on global links. This reduces congestion-induced slowdowns and improves performance consistency at scale. 
%The performance gain is more noticeable for medium-sized transfers as, on Slingshot, for large transfers the congestion control and adaptive routing algorithms typically reach a steady state, reducing slowdowns induced by congestion~\cite{sc2020}.
%In the few cases where \Bine is not the most performing algorithm, standard binomial trees are but, as discussed in Sec.~\ref{sec:evaluation:lumi:bvb}, the performance gap remains below 10\%. %Such small performance gaps, especially for short vector sizes, are likely attributable to minor implementation details, as the differences correspond to only a few microseconds.

%\paragraph{\textbf{Reduce-Scatter}} 
%We also observe that for the largest vector size, \bine trees are outperformed by the pairwise algorithm~\cite{}.
%or by the \textit{non-blocking} algorithm, which simply calls \texttt{MPI\_Iallreduce} collective, in this case using its default algorithm. 

\begin{figure}[t]
  \centering
  \begin{subfigure}[t]{0.48\columnwidth}
    \raisebox{1.5mm}{\includegraphics[width=\linewidth]{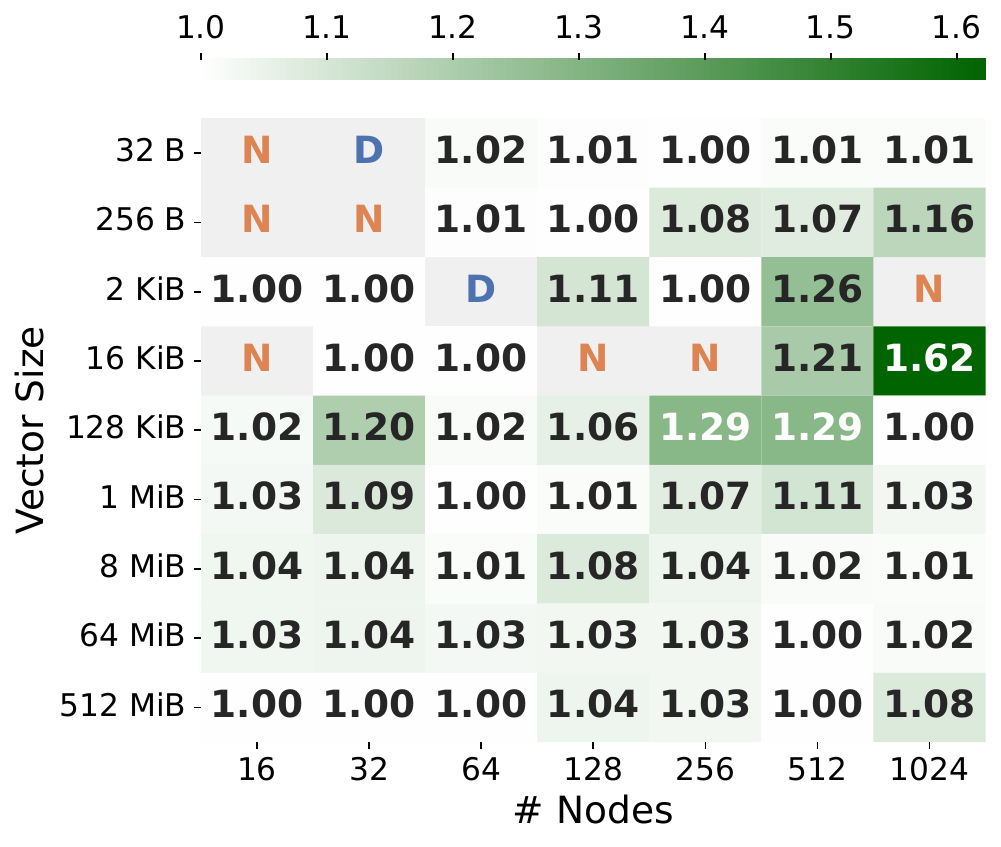}}%
    \caption{Allreduce}
    \label{fig:plot:lumi:allreduce}
  \end{subfigure}
  \begin{subfigure}[t]{0.48\columnwidth}
    \includegraphics[width=\linewidth]{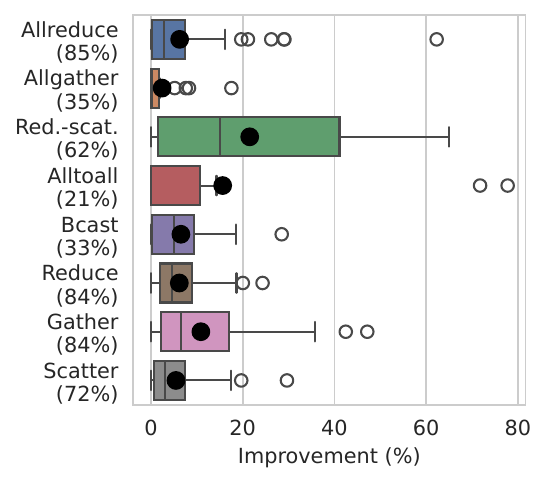}
    \caption{All collectives}
    \label{fig:plot:lumi:all}
  \end{subfigure}
  \caption{Comparison with state-of-the-art algorithms on LUMI.
(a): Each cell shows a letter indicating the best-performing algorithm (N = binomial, D = default), or, when \Bine is best, the performance ratio between \Bine and the next-best algorithm.
(b): For each collective, the number below the name indicates the percentage of tests where \Bine outperforms binomial trees. Box plots summarize the distribution of performance improvements in those cases.}
  \Description{Comparison of \Bine performance against state-of-the-art algorithms on LUMI.
Left: Each cell shows either a letter indicating the best-performing algorithm (N = binomial, D = default), or, when \Bine is best, the performance ratio between \Bine and the next-best algorithm.
Right: For each collective, the number below the name indicates the percentage of configurations where \Bine was the top performer. Box plots summarize the distribution of performance gains in those cases.}
  \label{fig:plot:lumi}
\end{figure}

\paragraph{\textbf{Other Collectives}}
For the remaining collectives, we summarize the corresponding heatmaps using boxplots in Fig.~\ref{fig:plot:lumi:all}, retaining only the configurations where \Bine outperforms all other algorithms. Below each collective name, we report the percentage of such cases.
For \reducescatter, we observe up to 65\% performance improvements for medium-sized vectors. As discussed in Sec.~\ref{sec:collectives:reducescatter_allgather} \reducescatter either sends non-contiguous blocks or permutes the buffer and sends contiguous blocks. 
%In the latter case, \Bine's better overlaps communication and computation. 
At large node counts, permuting the vector for contiguous data results in a 20\% performance drop for \Bine compared to binomial trees, as discussed in Sec.~\ref{sec:evaluation:lumi:bvb}.

For \allgather, \Bine was outperformed by the default Cray MPICH algorithm in 17\% of the cases. According to the Cray MPICH manual, the default algorithm leverages architecture-specific optimizations, which we suppose may include, for example, triggered operations. By disabling such optimizations, \Bine outperformed the state of the art in an additional 18\% of cases, with maximum performance gains reaching 89\%. A similar trend was observed for \bcast, with \Bine becoming the top-performing algorithm in 63\% of cases when disabling architecture-specific optimizations. 
For \alltoall, \Bine is the best algorithm in 21\% of cases. On 512 and \num{1024} nodes with 128 KiB vectors, it outperforms all others by up to 78\%. Like Bruck's algorithm, \Bine is optimized for small vectors, favoring fewer communication steps at the cost of more data per step. Hence, it performs best at larger scales and with smaller vectors.

\subsection{Evaluation on Leonardo}
Leonardo~\cite{turisini2023leonardo} uses a Dragonfly+ network topology based on Infiniband HDR. It consists of 23 groups, each containing 180 nodes, organized as two-level fat trees. We used the \textit{Booster} partition, where each node is equipped with two 200 Gb/s dual-port NVIDIA Connect-X6 NICs, a single-socket 32-core Intel Xeon\textregistered 8358 processor, 512 GB of host memory, and four NVIDIA A100 GPUs.

\begin{table}[htpb]
\footnotesize
\centering
\caption{Comparison with Binomial Trees on Leonardo}
\label{tab:results:summary:leonardo:bvb}
\resizebox{.85\columnwidth}{!}{
\begin{tabular}{lccccc}
\toprule
\textbf{Coll.} & \textbf{\% Win} & \makecell{\textbf{Avg/Max}\\\textbf{Perf. Gain}} & \textbf{\% Loss} & \makecell{\textbf{Avg/Max}\\\textbf{Perf. Drop}} & \makecell{\textbf{Avg/Max}\\\textbf{Traffic Red.}}  \\
\midrule
Allreduce & 67\% & 11\%/46\% & 20\% & 10\%/13\% & 19\%/26\% \\
Allgather & 91\% & 9\%/54\% & 0\% & 0\%/0\% & 19\%/26\% \\
Red.-Scat. & 71\% & 4\%/12\% & 14\% & 6\%/6\% & 17\%/23\% \\
Alltoall & 79\% & 25\%/70\% & 0\% & 0\%/0\% & 15\%/15\% \\
Bcast & 94\% & 41\%/148\% & 0\% & 0\%/0\% & 89\%/92\% \\
Reduce & 44\% & 43\%/72\% & 33\% & 7\%/9\% & 13\%/19\% \\
Gather & 93\% & 23\%/71\% & 7\% & 12\%/12\% & 12\%/20\% \\
Scatter & 93\% & 22\%/63\% & 0\% & 0\%/0\% & 12\%/20\% \\

\bottomrule
\end{tabular}
}
\end{table}

\subsubsection{\textbf{Comparison with Binomial Trees}}\label{sec:evaluation:leonardo:bvb}
Table~\ref{tab:results:summary:leonardo:bvb} compares \Bine to binomial trees on configurations from \num{16} to \num{2048} nodes, spanning 3 to 20 groups. Since standard users on \leonardo are limited to jobs of at most \num{256} nodes, the experiments on \num{512}, \num{1024}, and \num{2048} nodes were conducted during a maintenance window, with support from CINECA, on an otherwise idle system. To reduce the test duration and minimize impact on system operations, we collected data for only \allreduce and \allgather on jobs larger than \num{256} nodes.

The results show that \Bine outperforms binomial trees in most scenarios. For half of the collectives, \Bine is the best-performing algorithm in over 90\% of the cases and matches the performance of binomial trees in the remaining ones.. On \bcast we see higher advantages than on LUMI, especially for vectors smaller than 128 KiB, where performance gain where of 4\% on \lumi but of 42\% on \leonardo. This can be attributed to MPICH relying on distance-halving \bcast~\cite{mpich-bcast-binomial}, whereas Open MPI relies on distance-doubling \bcast~\cite{ompi-bcast-binomial,ompi-binomial}, which increases the number of bytes transmitted over global links, as we shown in Fig.~\ref{fig:bcast_motivation}.

\subsubsection{\textbf{Comparison with Other State-of-the-Art Algorithms}}\label{sec:evaluation:leonardo:sota}

\paragraph{\textbf{Allreduce}}

\begin{figure}[htpb]
  \centering
  \begin{subfigure}[t]{0.48\columnwidth}
    \raisebox{1.5mm}{\includegraphics[width=\linewidth]{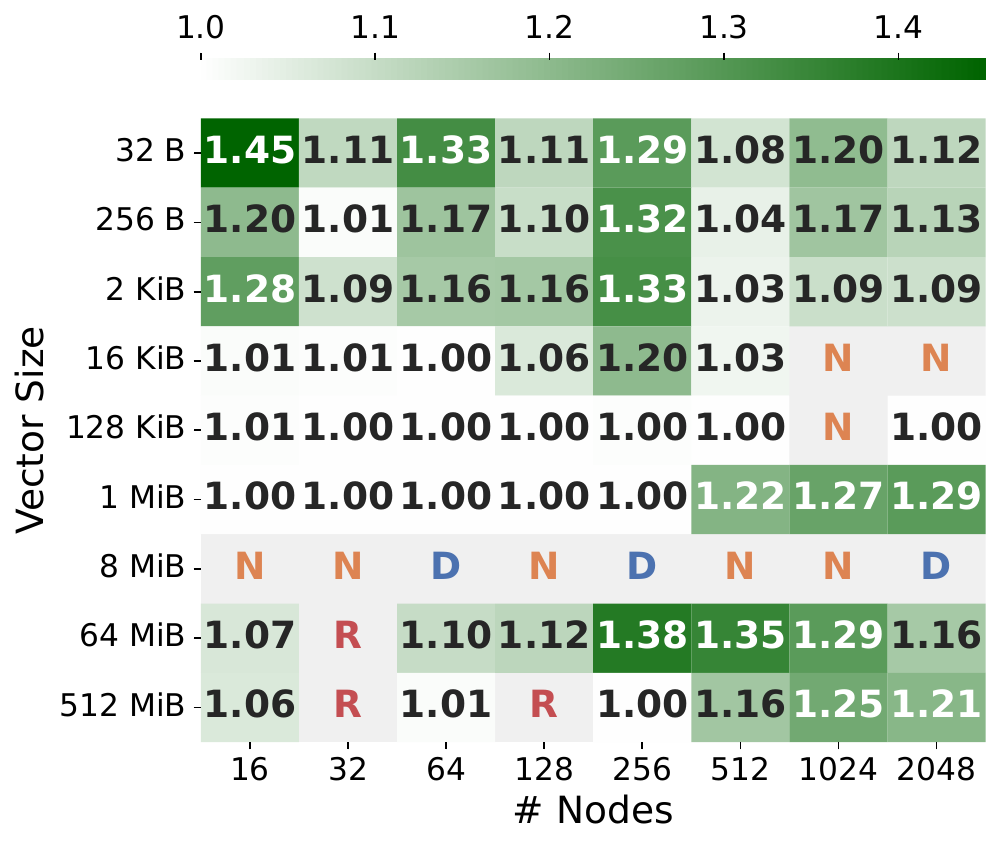}}%
    \caption{Allreduce}
    \label{fig:plot:leonardo:allreduce}
  \end{subfigure}
  \begin{subfigure}[t]{0.48\columnwidth}
    \includegraphics[width=\linewidth]{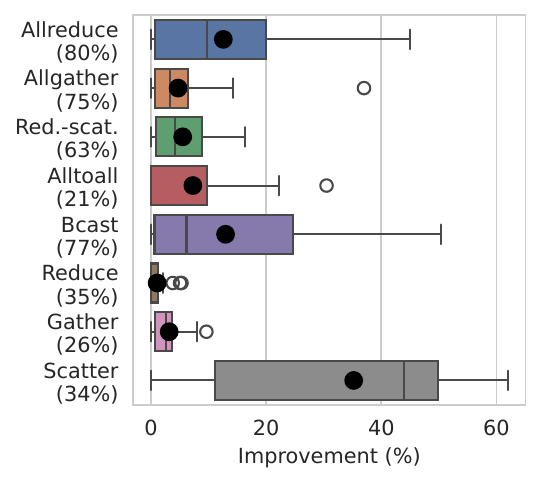}
    \caption{All collectives}
    \label{fig:plot:leonardo:all}
  \end{subfigure}
  \caption{Comparison with state-of-the-art algorithms on \leonardo
%Left: Each cell shows either a letter indicating the best-performing algorithm 
(N = binomial, D = default, R = ring).
%, or, when \Bine is best, the performance ratio between \Bine and the next-best algorithm.
%Right: For each collective, the number below the name indicates the percentage of configurations where \Bine was the top performer. Box plots summarize the distribution of performance gains in those cases.
}
  \Description{Comparison with state-of-the-art algorithms on \leonardo
%Left: Each cell shows either a letter indicating the best-performing algorithm 
(N = binomial, D = default, R = ring).
%, or, when \Bine is best, the performance ratio between \Bine and the next-best algorithm.
%Right: For each collective, the number below the name indicates the percentage of configurations where \Bine was the top performer. Box plots summarize the distribution of performance gains in those cases.
}
  \label{fig:plot:leonardo}
\end{figure}

%However, for larger node count the latency cost of it being linear in the number of nodes leads it to do not perform well anymore. We observe a performance drop of $~12\%$ for 8 MiB vectors. In this case, the best-performing state-of-the-art algorithm is the Rabenseifner reduce-scatter–allgather butterfly algorithm~\cite{}. We investigated this behavior further and found the same $\sim 12\%$ performance gap when comparing the Open MPI implementation of Rabenseifner’s algorithm with our porting of it on top of Open MPI. The only difference between the two lies in how buffers are copied: while we use a single memcpy, Open MPI 4.1.7 uses the \texttt{opal\_datatype\_copy\_content\_same\_ddt} function, which issues multiple smaller \texttt{memcpy} calls~\cite{openmpi-copy}. We measured that indeed, for an 8 MiB vector, 
%and we can reasonably assume this is the reason for the measured gap.
%We verified on $128–256$ nodes that, when fragmenting the memcpy as in Open MPI, the performance ratio becomes close to 1.0. Unfortunately, the $512-2048$ node runs were executed with support from CINECA during a maintenance window, and we were unable to re-run them. For consistency, we therefore report results using our implementation with a single memcpy, which leads to a higher performance drop.
We compare \allreduce performance with state-of-the-art algorithms in Fig.~\ref{fig:plot:leonardo:allreduce}. \Bine emerges as the best-performing algorithm in 67\% of the configurations, achieving performance improvements of up to $1.45\times$. For large vectors at small node counts, the ring algorithm is usually more effective. We verified that part of \Bine’s performance gains can be attributed to its use of segmentation during \allreduce, which enables better overlap between communication and computation. To isolate segmentation effects, we implemented a version of \Bine \allreduce without segmentation. 

Without segmentation, the ring algorithm outperforms \Bine for 512 MiB messages on \num{256} and \num{512} nodes. This is expected, as the ring algorithm inherently splits the data into smaller chunks than both \Bine and butterfly algorithms, which enhances overlap. To have a fair comparison with the ring algorithm, we present \bine's results with segmentation enabled. Still, except for these two cases, \Bine outperforms all other algorithms even without segmentation. 

\paragraph{\textbf{Other Collectives}}
We summarize the results in Fig.~\ref{fig:plot:leonardo:all}. For \allgather on \num{2048} nodes, vectors smaller than 64 MiB benefit from the remapping approach described in Sec.\ref{sec:collectives:allreduce}, achieving $2\times$ performance improvements compared to algorithms like \textit{Swing}~\cite{swing} or \textit{sparbit}~\cite{sparbit,ompi-allgather-sparbit}, which send non-contiguous data and introduce significant overhead. For \reducescatter, \Bine is the top-performing algorithm in 63\% of the cases. For \alltoall, results are similar to those on LUMI, with \Bine being the best algorithm on small vectors. 
%Performance improvements are more limited compared to LUMI, as \alltoall was run up to 256 nodes on Leonardo but up to \num{1024} nodes on LUMI. 
Finally, \Bine achieves higher performance on \bcast compared to LUMI, consistently with our findings in Sec.~\ref{sec:evaluation:leonardo:bvb}. %Last, we observe an average 33\% performance improvement on \scatter. This mostly happens for small vectors, and 

\subsection{Evaluation on MareNostrum 5}
\mn~\cite{marenostrum} uses a 2:1 oversubscribed fat-tree network topology based on InfiniBand NDR200. Each full-bandwidth sub-tree consists of 160 nodes. We utilized the ACC partition, where each node is equipped with $2\times$ Intel Sapphire Rapids 8460Y+ CPUs, 512 GB of DDR5 memory, and four 200 Gb/s NICs. Additionally, each node has 4x Nvidia Hopper GPUs with 64 GB of HBM2 memory.

\subsubsection{\textbf{Comparison with Binomial Trees}}\label{sec:evaluation:mn5:bvb}
We ran experiments from \num{4} to \num{64} nodes (the maximum allowed), spanning between one and eight subtrees. We report the results of our analysis in Table~\ref{tab:results:summary:mn:bvb}. Consistent with our findings on \lumi and \leonardo, \bine trees outperform binomial tree algorithms in most cases, with performance improvements of up to 158\%. However, in a few cases (more evident for \gat and \scatter), \bine increases (on average) the traffic on global links compared to binomial trees. We observed this mostly for small node count, consistently with what we discussed in Sec.~\ref{sec:bine}.

\begin{table}[ht]
\footnotesize
\centering
\caption{Comparison with Binomial Trees on \mn}
\label{tab:results:summary:mn:bvb}
\resizebox{.85\columnwidth}{!}{
\begin{tabular}{lccccc}
\toprule
\textbf{Coll.} & \textbf{\% Win} & \makecell{\textbf{Avg/Max}\\\textbf{Perf. Gain}} & \textbf{\% Loss} & \makecell{\textbf{Avg/Max}\\\textbf{Perf. Drop}} & \makecell{\textbf{Avg/Max}\\\textbf{Traffic Red.}}  \\
\midrule
Allreduce & 80\% & 9\%/46\% & 2\% & 6\%/6\% & 2\%/11\% \\
Allgather & 55\% & 10\%/42\% & 24\% & 12\%/17\% & 0\%/11\% \\
Red.-Scat. & 74\% & 15\%/78\% & 2\% & 6\%/6\% & 0\%/11\% \\
Alltoall & 88\% & 26\%/62\% & 10\% & 11\%/12\% & 19\%/33\% \\
Bcast & 98\% & 32\%/158\% & 0\% & 0\%/0\% & 49\%/86\% \\
Reduce & 51\% & 31\%/80\% & 22\% & 19\%/40\% & 2\%/7\% \\
Gather & 69\% & 25\%/76\% & 2\% & 8\%/8\% & -8\%/18\% \\
Scatter & 95\% & 36\%/140\% & 5\% & 7\%/8\% & -8\%/18\% \\
\bottomrule
\end{tabular}
}
\end{table}

\subsubsection{\textbf{Comparison with Other State-of-the-Art Algorithms}}\label{sec:evaluation:mn5:sota}
\begin{figure}[t]
  \centering
  \begin{subfigure}[t]{0.48\columnwidth}
    \includegraphics[width=\linewidth]{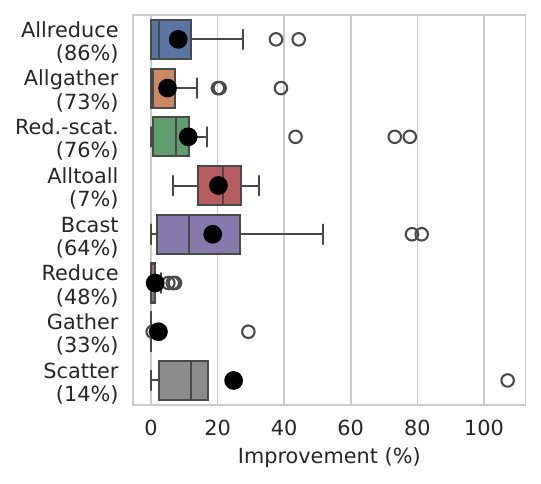}
    \caption{\mn}
    \label{fig:plot:mn:all}
  \end{subfigure}  
  \begin{subfigure}[t]{0.49\columnwidth}
    \includegraphics[width=\linewidth]{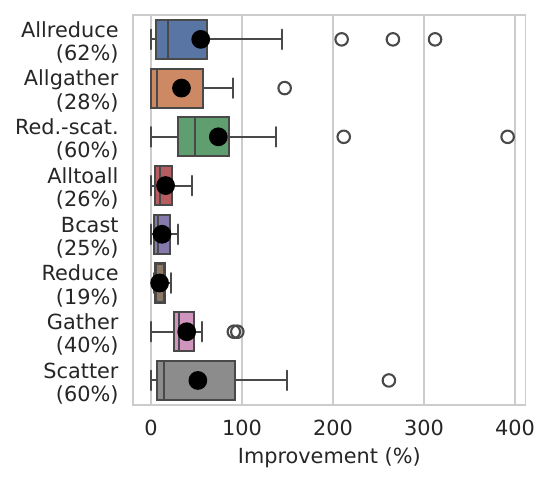}
    \caption{\fugaku}
    \label{fig:plot:fugaku:all}
  \end{subfigure}
  \caption{Comparison with state-of-the-art algorithms. %For each collective, the number below the name indicates the percentage of configurations where \Bine was the top performer. Box plots summarize the distribution of performance gains in those cases.
  }
   \Description{Comparison with state-of-the-art algorithms. %For each collective, the number below the name indicates the percentage of configurations where \Bine was the top performer. Box plots summarize the distribution of performance gains in those cases.
  }
  \label{fig:plot:mn}
\end{figure}

Fig.~\ref{fig:plot:mn:all} summarizes the comparison between \Bine and the best state-of-the-art algorithms. The results are consistent with our observations on the other platforms. For \alltoall, \gat, and \scatter, \Bine is the best-performing algorithm in fewer cases. This is primarily due to the smaller scale of the experiments (up to 64 nodes), compared to \num{1024} or \num{256} nodes on \lumi and \leonardo. At these scales, linear algorithms often outperform logarithmic ones, especially for large vectors. Conversely, for vectors smaller than 1 MiB, \Bine outperformed the other algorithms by up to $2\times$ on 64 nodes.

\subsection{Evaluation on Fugaku}\label{sec:evaluation:fugaku}
Fugaku~\cite{fugaku} is based on the Tofu Interconnect D~\cite{tofu} and uses a 6D toroidal network. In practice, each job can be allocated a 3D sub-torus. The system is composed of almost 159,000 nodes, each equipped with a Fujitsu A64FX CPU, 32 GiB of HBM memory, and six TNIs (i.e., NICs), each capable of injecting 54.4 Gb/s.

\subsubsection{\textbf{\Bine Optimization for Torus Networks}}
So far, we have not made any assumptions about the network topology. However, we can extend \bine trees to better exploit communication locality on torus networks. Instead of considering a flat rank space, we treat ranks as coordinates in a multidimensional space, reducing distance between communicating ranks, as typical in algorithms optimized for toroidal networks~\cite{bucket1,swing}. 
%Each rank sends message over one dimension at a time. For example, in a \reduce collective, in the first step, rank 0 (with negabinary coordinates $(00,00)$) receives from the rank whose \textit{first} coordinate differs by the least significant bit (i.e., $(01,00)$, rank 4). In the second step, it receives from the rank whose \textit{second} coordinate differs by the least significant bit (i.e., $(00,01)$, rank 1), rather than receiving from a rank $q$ such that $rank2nb(q) = 0011_{-2}$. This process continues for $\log_2 p$ steps, cycling through the coordinates. By doing so, distance between communicating ranks is shorter. On a torus network, minimizing communication distance is particularly important, as longer distances between communicating ranks increase contention on the links. For example, if each rank communicates with a peer at distance $d$, one or more links will be shared by $d$ communications, effectively reducing the injection bandwidth available to each node by a factor of $d$.
Since each node has six NICs, we divide the vector used in the collective into six parts. Then, we execute six collectives in parallel, each operating on a different part of the vector, using a separate NIC, and transmitting in a different direction of the 6D torus. To ensure that each collective uses a different NIC, we implemented the \bine algorithms directly on top of the low-level uTofu library~\cite{utofu}.
\ifshowextra
We provide a more detailed description of the design and the implementation of \bine trees for toroidal networks in the extended version of the paper~\cite{bine_arxiv}.
\else
We provide a more detailed description of the design and the implementation of \bine trees for toroidal networks in Appendix~\ref{apdx:torus}.
\fi

\subsubsection{\textbf{Comparison with Binomial Trees}}\label{sec:evaluation:fugaku:bnb}
When comparing \Bine to binomial trees, we observed up to $40\times$ speedups. However, this is largely because these binomial algorithms, based on Open MPI, on which Fujitsu MPI is based, are not optimized for toroidal networks. 

\subsubsection{\textbf{Comparison with Other State-of-the-Art Algorithms}}\label{sec:evaluation:fugaku:sota}
As with other systems, we compare both with the default algorithm selected by Fujitsu MPI and with all other available algorithms by explicitly forcing their selection. 
%Additionally, following the Fujitsu MPI manual's recommendations, we set the \texttt{coll\_tuned\_prealloc\_size} to a sufficiently large value for all collectives. This setting improves the performance of large vector collectives by pre-allocating a scratch buffer, thus avoiding repeated allocation during execution. To ensure a fair comparison, we implemented a similar pre-allocation approach for \bine algorithms. 
We performed our analysis on \texttt{2x2x2}, \texttt{4x4x4}, \texttt{8x8x8}, \texttt{64x64}, and \texttt{32x256} nodes jobs. 
We report results in Fig.~\ref{fig:plot:fugaku:all}. For \allreduce, \Bine trees outperformed state-of-the-art algorithms in 62\% of cases, especially for vectors larger than 2 KiB. The highest gains—up to $4.12\times$—were seen for 128 KiB–64 MiB vectors on \num{8192} (\texttt{32x256}) nodes. While Fugaku uses the torus-optimized Trinaryx algorithm~\cite{tofu,oai:ipsj.ixsq.nii.ac.jp:00214220,topotofu}, which builds three edge-disjoint spanning trees, its linear number of steps limits scalability compared to \Bine's logarithmic approach.

%We summarize the results for the other collectives in Fig.~\ref{fig:plot:fugaku:all}.
%For the \allreduce, \bine trees outperformed the available state-of-the-art algorithms in 67\% of the cases, mostly for vectors larger than 2 KiB. We observed the highest performance improvements, ranging between $1.62\times$ and $4.12\times$, for vectors in the range 128 KiB - 63 MiB on a \texttt{32x256} (i.e., \num{8192}) nodes torus. On Fugaku, the \allreduce collective uses the \textit{Trinaryx} algorithm, implementing \allreduce as a pipelined \reduce followed by a \bcast. This algorithm builds three edge-disjoint spanning trees, ensuring that each link is used by only one rank. However, it performs a number of steps linear with the number of nodes per dimension, differently from \bine trees that performs a logarithmic number of steps, enabling them to outperform \textit{Trinaryx} across most vector sizes and node counts.
%We also observe that for small vectors, \bine trees are slightly outperformed by standard binomial trees. In practice, this corresponds to differences of just a few microseconds, which we attribute to implementation inefficiencies in our current prototype, particularly evident at small message sizes.
For \allreduce, \reducescatter, and \scatter, \bine trees were the top-performing algorithm in over 60\% of the tests, with performance gains of up to $5\times$ for \reducescatter. In contrast, for \bcast and \reduce, Fujitsu MPI includes algorithms that are optimal on multiported torus networks, leaving little room for improvement. Nonetheless, the fact that \bine remains competitive, even against these highly optimized vendor implementations, underscores its strong generalization capability and broad effectiveness.

It is worth noting that Fugaku exhibits the highest performance gains among the four analyzed systems. These gains stem from Bine’s ability to reduce traffic on oversubscribed links. In the other three topologies, oversubscription primarily affects global links. In contrast, on a torus, all links can be considered oversubscribed. 
%For example, in an 8-node 1D torus executing a distance-doubling reduce-scatter, the second step involves communications between ranks \texttt{0<->2}, \texttt{1<->3}, \texttt{4<->6}, and \texttt{5<->7}. The link between nodes \texttt{1} and \texttt{2} is used by both the \texttt{0<->2} and \texttt{1<->3} communications, creating contention. Bine, on the other hand, schedules communications as \texttt{0<->7}, \texttt{1<->2}, \texttt{3<->4}, and \texttt{5<->6}, ensuring that each link is used by only one message. This strategy can double the performance of individual steps, and the improvements accumulate over the course of the operation. 
From another perspective, torus networks have the lowest bisection bandwidth among the topologies considered, and thus benefit the most from Bine’s enhanced locality.

%For \allreduce, \reducescatter, and \scatter, \bine trees emerge as the best-performing algorithm in over 60\% of the scenarios, with performance gains of up to $5\times$ for \reducescatter. These results further demonstrate the effectiveness of \bine trees across a variety of network topologies. For \bcast and \reduce, Fujitsu MPI provides algorithms that are bandwidth-optimal on multiported torus networks, leaving limited room for improvement. Still, the competitiveness of \bine, even against highly tuned vendor implementations, highlights its strong generalization potential and effectiveness.
%However, compared to the other three systems, \bine trees are the top-performing choice in fewer cases on Fugaku. This is due to both algorithmic and implementation-specific factors. Fujitsu MPI includes several algorithms that are explicitly designed and finely tuned for multi-ported toroidal networks. In addition to the \allreduce algorithm discussed earlier, this applies to both \bcast and \reduce, which use a similar strategy. 

%In addition, our prototype does not yet fully exploit the capabilities of uTofu. For instance, it does not currently use triggered operations, which could improve the performance of small-vector collectives. Furthermore, for small messages, Fugaku offloads the execution and progress of \allreduce, \reduce, and \bcast to the TNIs, significantly lowering runtime and making further improvements more challenging.

\section{Discussion}\label{sec:discussion}
\subsection{{Impact of Processes per Node}}
%The previous tests were conducted with one process per node. Depending on the application, collectives can be executed with either one process per node or multiple processes per node. 
All experiments in Sec.~\ref{sec:evaluation} were conducted with one process per node. To study the impact of multiple processes per node, we ran all collectives on 64 nodes on LUMI with either one or four processes per node (LUMI has four NICs per node). Performance was largely consistent, though in a few cases \Bine achieved larger gains with four processes per node; for example, the 1 MiB \reducescatter gain increased from 59\% to 84\%. This is likely due to the higher traffic generated by each node when using multiple processes, which better highlights \Bine’s global traffic reduction benefits. We observed similar trends on \leonardo and \mn.

\subsection{{Multi-GPU Evaluation}}
We implemented a GPU-aware version of the \bine \allreduce algorithm on top of MPI. This was motivated by two main goals. First, it allows us to evaluate the performance of \bine on multi-GPU systems and directly compare it against NCCL's algorithms. Second, it enables the evaluation of a hierarchical implementation of \bine. We optimized our design and implementation for large vectors (i.e., implementing the \allreduce as a \reducescatter followed by an \allgather), and for the \mn and \leonardo architecture.

Each \mn (and \leonardo) node has four GPUs, fully connected to one another. Our hierarchical \bine implementation exploits this topology by first performing an intra-node \reducescatter, where each GPU exchanges data concurrently with the other three GPUs on the same node. Next, an inter-node \bine \allreduce is executed among ranks with the same GPU local identifier, operating on the buffer segment received during the \reducescatter phase. Finally, an intra-node \allgather reconstructs the complete result, following a similar pattern to the initial \reducescatter. As for NCCL, vectors are aggregated on the GPUs.

We evaluated \Bine against both Open MPI and NCCL (v2.20.5 on \mn and v2.22.3 on \leonardo). 
%Open MPI showed over $10\times$ lower performance than both \Bine and NCCL, mainly due to its host-based vector aggregation, causing multiple host-device memory transfers~\cite{gpugpuinterconnect}. 
%For vectors smaller than 4 MiB, \Bine was outperformed by NCCL, with the double binary tree algorithm being the best performer~\cite{nccl-dbt}, achieving up to $2\times$ higher performance. However, for
On \mn, NCCL outperforms MPI algorithms in most cases. However, our MPI implementation of \Bine surpasses NCCL’s best-performing algorithm across all configurations from 16 to 256 GPUs, for vector sizes larger than 4 MiB. On average, \Bine achieves a 5\% improvement, with gains reaching up to 24\% on 256 GPUs. On \leonardo, on 512 GPUs \Bine improves performance by 15\% compared to MPI and remains within 7\% of NCCL.

%This is because binary tree algorithms degrade at higher node counts and medium message sizes~\cite{9499917}, while logarithmic algorithms like \Bine perform better. 

%crucial to minimize global link traffic, which enhances the effectiveness of \Bine's optimizations. It is important to note that we did not specifically exploit the fact that multiple processes share the same node; the same algorithm and implementation were used in both cases.

%\begin{figure}
%    \centering
%    \includegraphics[width=\linewidth]{plots/lumi_hm_tpn/1,4.pdf}
%    \caption{Performance gain of \bine for one and four processes per node, on 64 nodes.}
%    \label{fig:plot:lumi:ppn}
%\end{figure}

\section{Related Work}
\hspace{\parindent} \textbf{Locality-Aware and Hierarchical Algorithm} Sevaral locality-aware, and hierarchical or multi-leader algorithms~\cite{10.1145/3555819.3555825,6061150, mleader,hidayetoglu2024hicclhierarchicalcollectivecommunication} have been proposed. These algorithms, however, either require knowledge of the underlying topology~\cite{6061150} or make assumptions about the partitioning of ranks across groups~\cite{10.1145/3555819.3555825}. %Some algorithms, like \textit{sparbit}~\cite{sparbit}, similar to \bine, do not make assumptions about the underlying topology or rank partitioning. However, \textit{sparbit} relies on standard binomial trees, and, as we have shown, \bine outperforms it in up to 90\% of the tested cases, reducing the number of bytes sent over global links by as much as 26\%.
%However, these algorithms still require strong assumptions about the mapping and partitioning of ranks to nodes, which is not necessary for \bine trees. 
On the other hand, \bine trees are as generic as binomial trees and, as shown in Sec.~\ref{sec:discussion}, can be effectively integrated into hierarchical algorithms.

\textbf{Automatic Algorithm Synthesis} Some methods aim to automatically determine the optimal collective algorithm based on network topology and collective specifications~\cite{shah2022taccl, 10.1145/3437801.3441620, won2024tacostopologyawarecollectivealgorithm}. However, their complexity increases exponentially as the system scales. For example, computing a solution for 128 nodes can take up to 11 hours~\cite{shah2022taccl}, and a new solution may need to be recalculated for each allocation, which makes these approaches not practical on shared supercomputers, where users do not have control over allocations.

\textbf{Topology-Optimized Algorithms} Researchers developed collective algorithms tailored to specific network topologies~\cite{10.1145/1088149.1088183, 9644896, 10.1145/3524059.3532380}. In contrast, \Bine trees are versatile and topology-agnostic and consistently enhance performance across a variety of topologies.

\section{Conclusions}
In this paper, we introduced \Bine trees, a novel approach to constructing binomial trees that reduces the distance between communicating ranks and, as a result, reduces traffic on global links and improves performance. We designed new algorithms for eight different collective operations, outperforming state-of-the-art algorithms, including traditional binomial trees and topology-specific methods. We observed performance improvements of up to $5\times$ on \num{8192} nodes and reductions in global link traffic of up to 33\%, highlighting the potential of \Bine to deliver significant gains in performance and efficiency across different network topologies. 

The main strength of \Bine trees lies in their generality, being as generic as standard binomial trees and not relying on specific assumptions about the underlying network topology or job allocation. This makes it well-suited for large-scale, shared supercomputing environments. In conclusion, thanks to their broad applicability and ability to reduce network traffic across global links, \Bine trees represent a significant step toward more efficient, scalable, and flexible collective algorithms. They can effectively replace binomial trees and butterfly-based approaches, especially at large scales.

\begin{acks}

We thank Daniele Di Bari (CINECA) for support with large-scale tests on Leonardo, Kim McMahon (HPE Cray) for assistance with Cray MPICH collective algorithm selection, Kurt Lust for information on node-to-group mapping on LUMI, and Faveo Hörold (ETH Zürich and RIKEN) for help with preliminary tests on Fugaku. This work is supported by the European Union’s Horizon Europe under grant 101175702 (NET4EXA), by Sapienza University Grants ADAGIO and D2QNeT (\textit{Bando per la ricerca di Ateneo} 2023 and 2024), and by the European Research Council (ERC) under the European Union’s Horizon 2020 research and innovation program (grant PSAP, No. 101002047). Daniele De Sensi is a member of GNCS-INdAM. The authors used ChatGPT for editing; all ideas, content, and conclusions are their own.
\end{acks} 
%%
%% The acknowledgments section is defined using the "acks" environment
%% (and NOT an unnumbered section). This ensures the proper
%% identification of the section in the article metadata, and the
%% consistent spelling of the heading.
%\begin{acks}
% Faveo
% Daniele Di Bari, Kim McMahon
%\end{acks}

%%
%% The next two lines define the bibliography style to be used, and
%% the bibliography file.
\bibliographystyle{ACM-Reference-Format}
\bibliography{main}

%%
%% If your work has an appendix, this is the place to put it.
\ifshowextra
\else
\newpage
\clearpage
\newpage
\begin{appendix}

\section{Detailed Description of Distance-Doubling \Bine Trees}\label{apdx:dist-doubling-trees}    
Since we want to preserve the $33\%$ modular distance reduction compared to distance-doubling binomial trees, in distance-doubling \bine trees, ranks communicate in the opposite order compared to distance-halving butterflies. Specifically, after receiving data at step $i$ (as described in Sec.~\ref{sec:bine:doubling}), at each subsequent step $j>i$ a rank $r$ sends data to a rank $q$ determined as:

\begin{equation}\label{eq:apdx:distdoub}
q = \begin{cases}  
    \Big(r + \sum_{k=0}^{j}(-2)^k \Big) \bmod p, & \text{if } r \text{ is even}, \\
    \Big(r - \sum_{k=0}^{j}(-2)^k \Big) \bmod p, & \text{if } r \text{ is odd}.
\end{cases}
\end{equation} 

The modular distance between two communicating ranks is thus $\sum_{k=0}^{j}(-2)^k$, implying a $33\%$ reduction in distance compared to standard distance-doubling trees (for the reasons discussed in Sec.~\ref{sec:bine:halving:advantage:theory}). Note that $\sum_{k=0}^{j}(-2)^k = \tfrac{1}{3} - \tfrac{-2^{j+1}}{3}$ is always odd, meaning that even ranks always communicate directly with odd ranks, and vice versa.

We now analyze when the data sent by the root rank $r$ reaches a given rank $q$. For simplicity, we consider a broadcast with root $r=0$, but the same reasoning applies to other collectives and to arbitrary roots $r \neq 0$ (in that case, the tree is simply rotated by $r$ positions). At step $j$, the root $r=0$ sends data to rank 
\[
q = \sum_{k=0}^{j}(-2)^k \bmod p.
\]
$q$ is odd, and when $q$ later forwards the data at some step $c>j$, it sends it to:
\[
t = q - \sum_{k=0}^{c}(-2)^k \bmod p = \sum_{k=0}^{j}(-2)^k - \sum_{k=0}^{c}(-2)^k \bmod p.
\]
i.e., the sign between each summation alternates from positive to negative and vice-versa.

To generalize, consider the path through which data from root $0$ reaches a rank $q$ after $h$ steps $s_0 < s_1 < \dots < s_{h-1}$.  
If $q$ is odd, since communication alternates between even and odd ranks, $h$ must be odd, and:
\[
q = \sum_{i=0}^{s_0} (-2)^i - \sum_{i=0}^{s_1} (-2)^i + \sum_{i=0}^{s_2} (-2)^i - \dots + \sum_{i=0}^{s_{h-1}} (-2)^i \mod p.
\]
Simplifying the summations gives:
\begin{equation}\label{apdx:eq:negabinary}
q = \sum_{i=0}^{s_0} (-2)^i + \sum_{i=s_1+1}^{s_2} (-2)^i + \dots + \sum_{i=s_{h-2}+1}^{s_{h-1}} (-2)^i \mod p.
\end{equation}

The right-hand side corresponds to a negabinary representation, whose least significant bit is $1$, as illustrated in Fig.~\ref{fig:binary_string_alternating_s0} for a 12-bit string with $s_0=1$, $s_1=5$, and $s_2=8$.

\begin{figure}[h!]
\centering
\begin{tabular}{|c|c|c|c|c|c|c|c|c|c|c|c|c|}
    \multicolumn{3}{c}{} & \multicolumn{1}{c}{$s_2$} & \multicolumn{2}{c}{}  & \multicolumn{1}{c}{$s_1$} & \multicolumn{3}{c}{}  & \multicolumn{1}{c}{$s_0$} & \multicolumn{1}{c}{} \\
    \multicolumn{3}{c}{} & \multicolumn{1}{c}{$\underset{\downarrow}{\hphantom{1\dots1}}$} & \multicolumn{2}{c}{}  & \multicolumn{1}{c}{$\underset{\downarrow}{\hphantom{1\dots1}}$} & \multicolumn{3}{c}{}  & \multicolumn{1}{c}{$\underset{\downarrow}{\hphantom{1\dots1}}$} & \multicolumn{1}{c}{} \\
    \hline
    0 & 0 & 0 & 1 & 1 & 1 & 0 & 0 & 0 & 0 & 1 & 1 \\
    \hline
\end{tabular}
\caption{Negabinary representation corresponding to Eq.~\ref{apdx:eq:negabinary} with $s_0=1$, $s_1=5$, and $s_2=8$.}
\label{fig:binary_string_alternating_s0}
\end{figure}

In this representation, the leftmost and rightmost positions of each contiguous sequence of $1$s correspond to the values of the $s_i$. Thus, if $q$ is odd and its negabinary encoding matches Fig.~\ref{fig:binary_string_alternating_s0}, the data from the root reached $q$ through step $s_0=1$, then $s_1=5$, and finally $s_2=8$.  

To isolate the positions of the $s_i$ (i.e., to create a binary string with 1s only at those positions, e.g., $000100100010$ for the example just described), we define $h(q,p)$ as the negabinary string in Fig.~\ref{fig:binary_string_alternating_s0}, and compute:
\[
\nu(q,p) = h(q,p) \oplus (h(q,p) \gg 1).
\]
In essence, by inspecting the bits of $\nu(q,p)$, we can reconstruct exactly how the data propagated from the root to rank $q$.

Rank $q$ receives the data at step $s_{h-1}$, corresponding to the most significant bit set to $1$ in $\nu(q,p)$. At each subsequent step $j$, it forwards the data to a rank whose $\nu$ differs only in bit $j$—that is, the destination rank has a $1$ in position $j$ and $0$s in all more significant positions, consistent with the algorithm’s design.

The same reasoning extends to even ranks. In that case, $h$ is even, and:
\[
q = \sum_{i=0}^{s_0} (-2)^i - \sum_{i=0}^{s_1} (-2)^i + \sum_{i=0}^{s_2} (-2)^i - \dots - \sum_{i=0}^{s_{h-1}} (-2)^i \mod p,
\]
which simplifies to:
\[
q = -\sum_{i=s_0+1}^{s_1} (-2)^i - \sum_{i=s_2+1}^{s_3} (-2)^i - \dots - \sum_{i=s_{h-2}+1}^{s_{h-1}} (-2)^i \mod p.
\]
Multiplying both sides by $-1$ yields:
\[
-q = \sum_{i=s_0+1}^{s_1} (-2)^i + \sum_{i=s_2+1}^{s_3} (-2)^i + \dots + \sum_{i=s_{h-2}+1}^{s_{h-1}} (-2)^i \mod p.
\]
Adding $p$ to both sides, we obtain:
\begin{equation}\label{apdx:eq:negabinary2}
p - q = \sum_{i=s_0+1}^{s_1} (-2)^i + \sum_{i=s_2+1}^{s_3} (-2)^i + \dots + \sum_{i=s_{h-2}+1}^{s_{h-1}} (-2)^i \mod p.
\end{equation}

The right-hand side again represents a negabinary number, but with its least significant bit equal to 0, as shown in Fig.~\ref{fig:binary_string_alternating_s1} for $s_0=1$, $s_1=5$, $s_2=8$, and $s_3=10$. For this reason, for even $q$ we define:
\[
\nu(q,p) = h(p-q,p) \oplus (h(p-q,p) \gg 1).
\]

\begin{figure}[htpb]
\centering
\begin{tabular}{|c|c|c|c|c|c|c|c|c|c|c|c|c|}
    \multicolumn{1}{c}{} & \multicolumn{1}{c}{$s_3$} & \multicolumn{1}{c}{} & \multicolumn{1}{c}{$s_2$} & \multicolumn{2}{c}{}  & \multicolumn{1}{c}{$s_1$} & \multicolumn{3}{c}{}  & \multicolumn{1}{c}{$s_0$} & \multicolumn{1}{c}{} \\
    \multicolumn{1}{c}{} & \multicolumn{1}{c}{$\underset{\downarrow}{\hphantom{1\dots1}}$} & \multicolumn{1}{c}{} & \multicolumn{1}{c}{$\underset{\downarrow}{\hphantom{1\dots1}}$} & \multicolumn{2}{c}{}  & \multicolumn{1}{c}{$\underset{\downarrow}{\hphantom{1\dots1}}$} & \multicolumn{3}{c}{}  & \multicolumn{1}{c}{$\underset{\downarrow}{\hphantom{1\dots1}}$} & \multicolumn{1}{c}{} \\
    \hline
    0 & 1 & 1 & 0 & 0 & 0 & 1 & 1 & 1 & 1 & 0 & 0 \\
    \hline
\end{tabular}
\caption{Negabinary representation corresponding to Eq.~\ref{apdx:eq:negabinary2}, with $s_0=1$, $s_1=5$, $s_2=8$, and $s_3=10$.}
\label{fig:binary_string_alternating_s1}
\end{figure}

\section{Comparison of Non-Contiguous Data Transfers Techniques}\label{apdx:noncontiguous}

\begin{figure}[H]
    \centering
    \includegraphics[width=\linewidth]{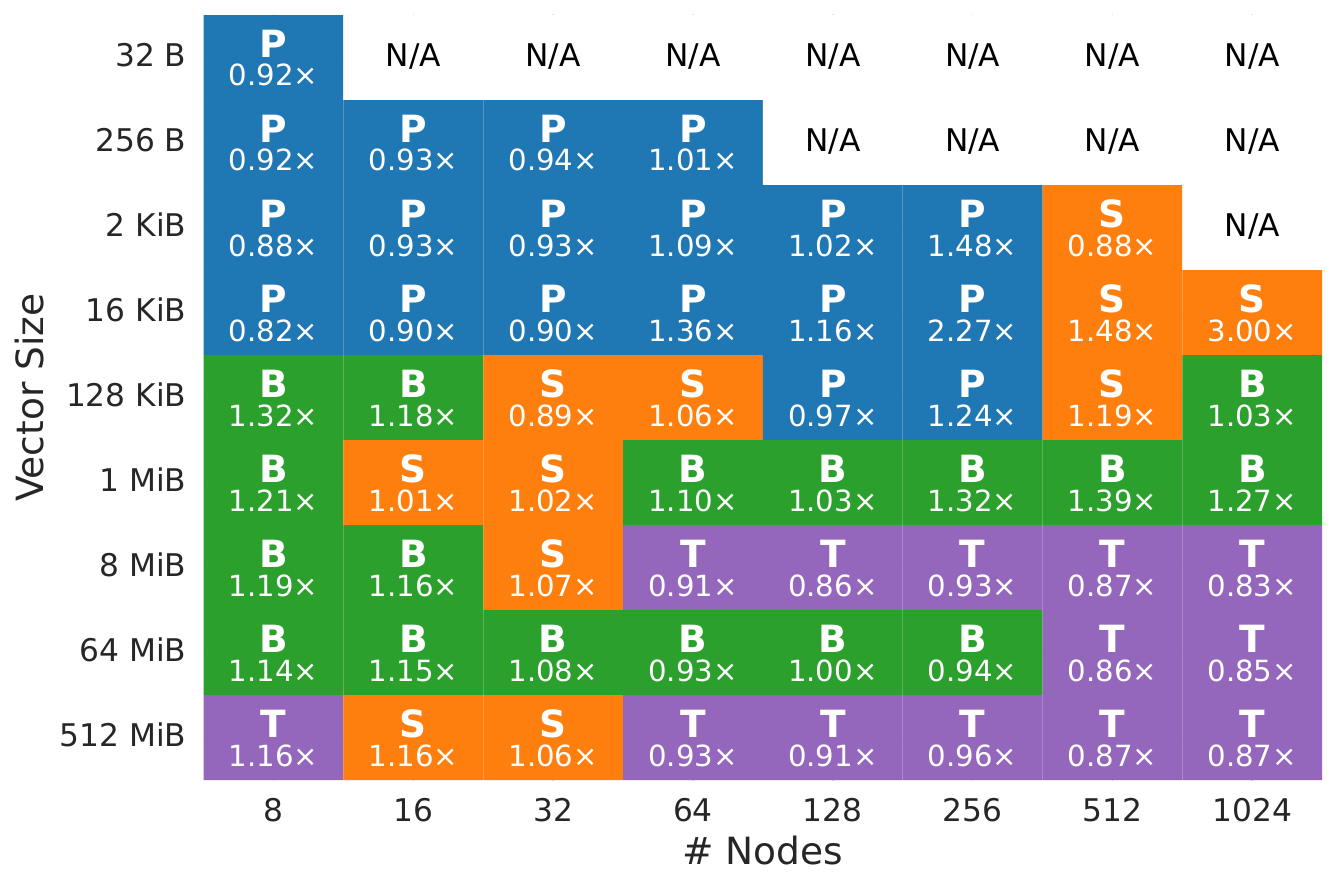}
    \caption{Comparison of techniques for handling the transmission of non-contiguous data (see Sec.~\ref{sec:bine:noncontiguous}). Results are obtained on LUMI for an \allgather collective operation. Each cell shows a letter corresponding to the best technique (B=\textbf{Block-by-block}, P=\textbf{Permute}, S=\textbf{Send}, T=\textbf{Two Transmissions}), and the performance gain compared to standard binomial butterflies.}
    \Description{Comparison of techniques for handling the transmission of non-contiguous data (see Sec.~\ref{sec:bine:noncontiguous}). Each cell shows a letter corresponding to the best technique (B=\textbf{Block-by-block}, P=\textbf{Permute}, S=\textbf{Send}, T=\textbf{Two Transmissions}), and the performance gain compared to standard binomial butterflies.}
    \label{fig:bine_noncontiguouos_variants}
\end{figure}

In Fig.~\ref{fig:bine_noncontiguouos_variants}, we compare the different techniques for handling the transmission of non-contiguous data (see Sec.~\ref{sec:bine:noncontiguous}) for an \allgather executed on \lumi. Each cell reports the best-performing technique (B=\textbf{Block-by-block}, P=\textbf{Permute}, S=\textbf{Send}, T=\textbf{Two Transmissions}) together with its performance gain relative to standard binomial butterflies. 
We first observe that, for small vectors, the \textit{permute} technique achieves the highest performance, with gains of up to 2.27$\times$ over binomial butterflies. Since the number of blocks, and thus the number of memory copies required for the permutation, grows with the number of nodes, the cost of permutation increases with system scale. As a result, for larger node counts, the \textit{send} technique becomes more effective, as it avoids buffer permutation and instead adds a final communication step to restore the correct order.

For larger vectors, the cost of permutation further increases because more data must be rearranged, making the \textit{block-by-block} technique more efficient. However, as the number of nodes grows, the number of transmissions also rises, and the best-performing approach becomes the \textit{two transmissions} technique. 
Finally, we note that the current implementation still offers opportunities for optimization. For instance, in the \textit{permute} variant, data permutation could be overlapped with communication by employing an additional buffer, in which data received during the previous step is permuted while data for the current step is transmitted.

\section{Managing Non-Power of Two Ranks}\label{apdx:nonp2}
If the number of ranks is not a power of $2$, different strategies can be adopted. The most straightforward approach follows techniques commonly used in binomial trees~\cite{mpich-bcast-binomial}. For clarity, consider a one-dimensional torus of size $p$ not equal to a power of two. Let $p' = 2^{\lfloor \log_2 p \rfloor}$. For tree-based collectives (e.g., a broadcast), one option is to first execute the collective among $p'$ ranks, and then have the remaining $p-p'$ ranks receive the data from $p-p'$ of the already-participating ranks. 

For butterfly-based collectives (e.g., reduce-scatter), a straightforward solution is to have the last $p-p'$ ranks forward their data to the first $p-p'$ ranks, then execute the reduce-scatter on these $p'$ ranks, and finally redistribute part of the resulting data back to the last $p-p'$ ranks. However, this doubles the total communication volume, since the first $p-p'$ ranks must first receive an entire vector from one of the last ranks before participating in the reduce-scatter.

\begin{figure}[htpb]
  \centering
  \includegraphics[width=\linewidth]{fig/bine_nonp2.pdf}
  \caption{\Bine trees constructed for a number of ranks that is not a power of $2$.}
  \label{fig:bine_nonp2}
\end{figure}

If $p$ is even, we can adopt a simpler approach that does not increase the total communication volume. In this case, the collective is executed as usual, but some ranks may be reached twice. This situation is illustrated in Fig.~\ref{fig:bine_nonp2} for collectives with $6$ and $10$ nodes. For simplicity, a tree is shown, but the same applies to a butterfly, which, as discussed in Sec.~\ref{sec:bine:butterfly}, can be seen as the superposition of multiple trees, each rooted at a different rank.

For example, in the $6$-node case, both rank~$4$ and rank~$5$ are reached twice. This would compromise correctness. For instance, in a reduce operation, the same contribution might be applied twice. To avoid this issue, it suffices to remove one of the two subtrees rooted at the rank that is reached twice. For example, for rank~$4$ we can either remove it as a leaf (and the children of rank~$5$), or remove the subtree rooted at rank~$4$ (which has rank~$5$ as a child).

To determine which of the two subtrees to remove, let us consider the $10$-node tree shown in Fig.~\ref{fig:bine_nonp2}. Each node is annotated with the step at which the data from the root reaches it (e.g., node $8$ is reached at step~$3$, since node $0$ sends the data to node $9$ at step~$2$, and then node $9$ forwards it to node $8$ at step~$3$). Since the subtree reached later always contains fewer nodes (which are also contained in the other subtree), we discard that one, as removing the larger subtree could compromise correctness. This is evident in the case of node $5$. We remove the subtree reached at step $3$ (Fig.~\ref{fig:bine_nonp2}~\raisebox{-0.2em}{\includegraphics[height=1em]{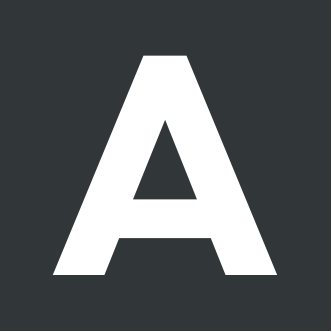}}), since removing the other one (Fig.~\ref{fig:bine_nonp2}~\raisebox{-0.2em}{\includegraphics[height=1em]{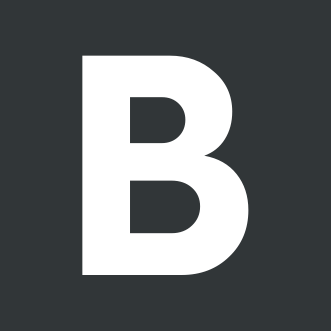}}) would disconnect a larger portion of the tree, leaving some nodes unreachable (e.g., nodes $6$ and $7$).

It is worth noting that this approach cannot be directly applied if $p$ is odd. Since communication always occurs in pairs, some ranks would inevitably attempt to communicate with a rank that is already engaged. In such cases, one must revert to the standard techniques discussed above.

\begin{figure*}[htpb]
  \centering
  \includegraphics[width=\linewidth]{fig/torus/bine_torus.pdf}
  \caption{Optimization of a distance-halving \bine tree for a 2D torus. For both \bine tree and torus-optimized \bine tree we show both the 2D torus on which the tree is built, and the corresponding tree.}
  \label{fig:bine_torus}
\end{figure*}

\section{Optimization for Torus Networks}\label{apdx:torus}
We now describe how to extend \Bine trees to better exploit communication locality on torus networks. For simplicity, we focus on trees; however, as discussed in Sec.~\ref{sec:bine:butterfly}, a butterfly can be viewed as a superposition of multiple trees, each rooted at a different rank. Therefore, the considerations in this section apply to both binomial trees and butterflies. We first assume that each torus dimension is a power of $2$, and then generalize to arbitrary dimension sizes.

\subsection{Selection of Communicating Partner}
We illustrate this in Fig.~\ref{fig:bine_torus}, comparing a standard \Bine tree with its torus-optimized counterpart on a $4 \times 4$ torus. Standard \Bine trees optimize for modular distance assuming a one-dimensional torus, which does not necessarily yield efficient communication paths in higher-dimensional topologies. For example, in the left part of Fig.~\ref{fig:bine_torus}, edge labels show the modular distance between communicating ranks as if they were arranged on a 1D torus. A sign indicates communication direction: a minus sign denotes communication to the left of the source rank. For instance, the modular distance between ranks $0$ and $15$ is computed as:
\[
d(0,15) = \min((0 - 15) \bmod 16,\ (15 - 0) \bmod 16) = \min(1, 15) = 1,
\]
which we label as $-1$ since the shorter path wraps leftward. However, on a 2D torus, the actual distance between ranks $0$ and $15$ is 2 hops (see Fig.~\ref{fig:bine_torus}~\raisebox{-0.2em}{\includegraphics[height=1em]{fig/torus/marker_a.pdf}}), traversing multiple links and potentially introducing additional congestion, as discussed in Sec.~\ref{sec:intro}.

To address this, we treat ranks as coordinates in a multidimensional space and apply the \Bine construction dimension by dimension, similar to other torus-optimized algorithms~\cite{bucket1,swing}. Communication partners are then chosen along single dimensions: instead of distances $-5,3,-1,1$, rank~$0$ exchanges with ranks at $(0,-1), (-1,0), (0,1), (1,0)$. Concretely, rank~$0$ at coordinates $(0,0)$ first communicates with $(0,3)$ (rank~3), then with $(3,0)$ (rank~12), then with $(0,1)$ (rank~1), and finally with $(1,0)$ (rank~4). As shown in Fig.~\ref{fig:bine_torus} \raisebox{-0.2em}{\includegraphics[height=1em]{fig/torus/marker_b.pdf}}, this approach reduces the number of crossed links, alleviating congestion and improving performance.

\subsection{Data Handling}\label{sec:bine:torus:data}
However, while this approach reduces communication distance, it complicates data management. To illustrate this issue and possible solutions, we consider a scatter operation implemented with a \Bine tree. Other collectives are handled similarly; for example, in a gather, data flows from the leaves to the root, and in an allgather, each rank uses its own tree. The root splits the data into \textit{``blocks''}, one per rank. Each node must forward to its children all the blocks belonging to the subtree rooted at that child. In standard distance-halving \bine trees, the descendants of any node are contiguous. For example, in Fig.~\ref{fig:bine_torus}, the subtree rooted at rank $11$ (highlighted by the dashed box) contains only consecutive ranks. As a result, when rank $0$ sends data to rank $11$, it transfers a set of contiguous memory blocks (i.e., the blocks in the range $[6, 13]$ Fig.~\ref{fig:bine_torus} \raisebox{-0.2em}{\includegraphics[height=1em]{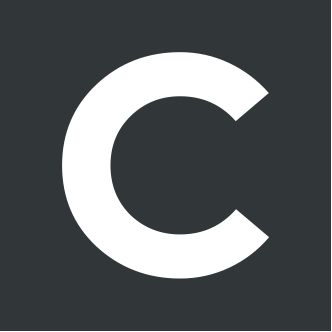}}).

In contrast, in torus-optimized \bine trees the descendants of a node are no longer contiguous, since ranks are arranged according to multidimensional coordinates. Returning to the scatter example, in the first step rank $0$ (with coordinates $(0,0)$) sends data to rank $3$ (with coordinates $(0,3)$). In this case, although those ranks are contiguous in the 2D torus (Fig.~\ref{fig:bine_torus} \raisebox{-0.2em}{\includegraphics[height=1em]{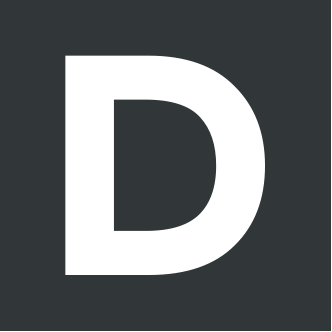}}), the blocks in the one-dimensional input vector are not (Fig.~\ref{fig:bine_torus} \raisebox{-0.2em}{\includegraphics[height=1em]{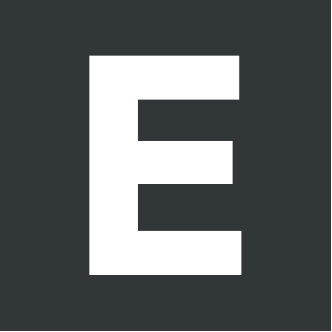}}). This increases data-handling complexity compared to the standard case, and can significantly hinder performance, especially for small- to medium-sized vectors~\cite{sparbit}, as we show in Sec.~\ref{sec:evaluation}.

An alternative is to permute the vector before transmission, ensuring that all subsequent sends by any rank involve contiguous data. This can be achieved by renumbering the nodes according to a \emph{DFS postorder} traversal (i.e., the order obtained by a depth-first search where a node is visited after all its children). For instance, in Fig.~\ref{fig:bine_torus}, the block originally at position $10$ (coordinates $(2,2)$) is moved to position $0$, the block at position $14$ (coordinates $(3,2)$) to position $1$, and so forth.

To avoid the initial permutation, ranks could send contiguous data as if the permutation had occurred. At the end of the collective, however, each rank would hold the wrong block; for instance, rank~$10$ (coordinates $(2,2)$) would receive the data intended for rank~$0$ (coordinates $(0,0)$). This can be corrected by adding an additional communication step at the end, where rank~$10$ sends its data to the correct rank (i.e., rank $0$), and other ranks perform analogous exchanges. 

Finally, some collectives are composed of multiple operations, such as an allreduce implemented as a reduce-scatter followed by an allgather. In these cases, even if the reduce-scatter places blocks in the wrong order, the subsequent allgather, which effectively inverts the reduce-scatter schedule, restores the blocks to the correct order.

\subsection{Non-Power of Two}
If one (or more) torus dimensions is not a power of $2$, we can rely on techniques commonly used in standard binomial trees and butterflies, as discussed in Appendix~\ref{apdx:nonp2}. On a multidimensional torus, however, beyond the challenges already mentioned, such techniques complicate the selection of the $p'$ ranks, particularly when trying to preserve short communication distances among participating ranks. As a consequence, several communication steps may require traversing long paths across multiple links. 

Moreover, running the collective on a subtorus with $p'$ ranks (whose dimensions are all powers of two) would disrupt the direct correspondence between torus coordinates and linear rank numbering, thereby complicating both process mapping and data management. When all torus dimensions are even, we instead adopt the approach described in Appendix~\ref{apdx:nonp2}, which eliminates duplicated subtrees. This strategy, illustrated in Fig.~\ref{fig:bine_torus_nonp2} for a $2\times6$ torus, does not require identifying a subtorus and does not add extra complexity compared to the power-of-two case.

\begin{figure}[htpb]
  \centering
  \includegraphics[width=.8\linewidth]{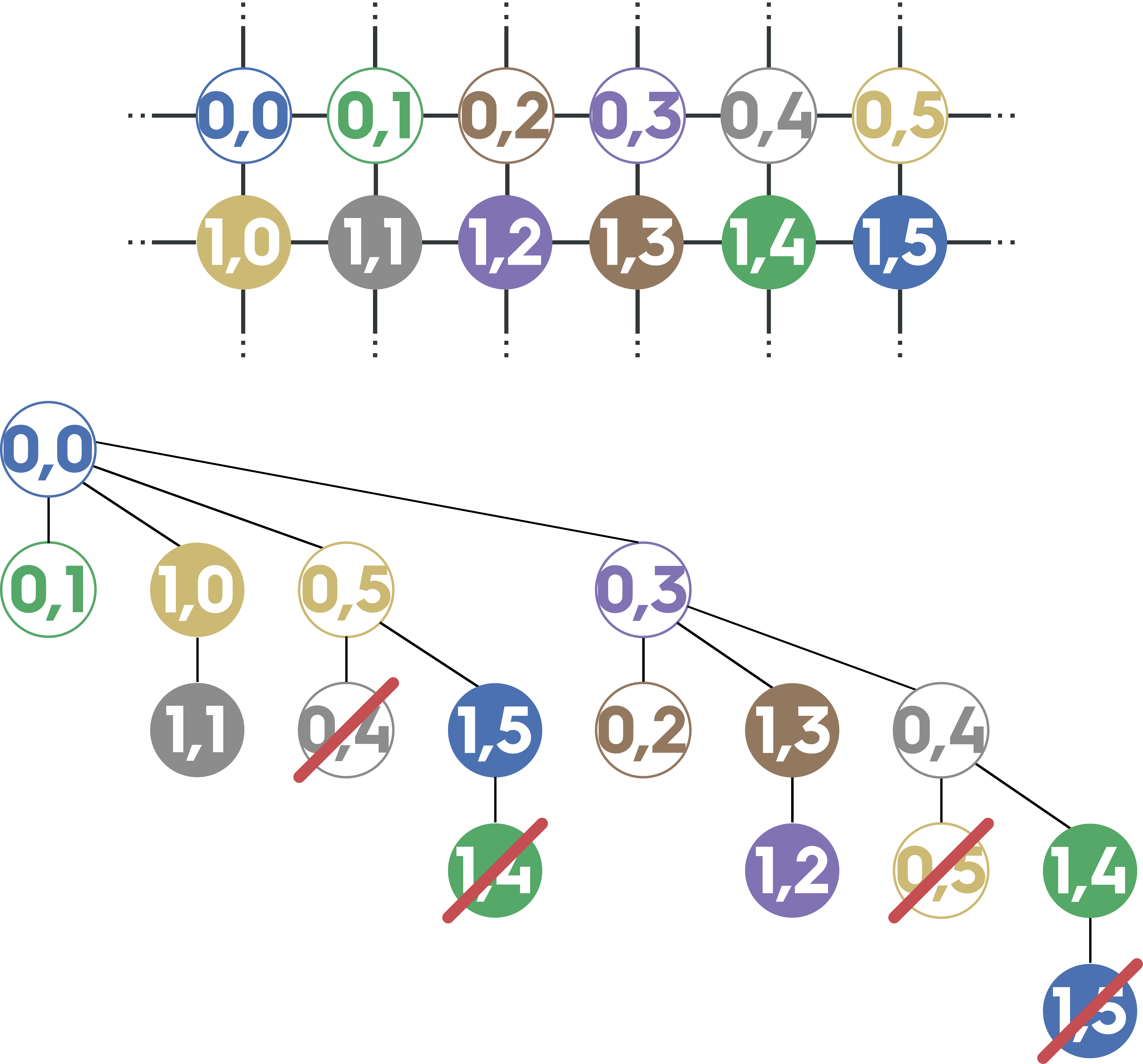}
  \caption{Torus-optimized \Bine trees for toroidal networks with dimensions that are not powers of $2$.}
  \label{fig:bine_torus_nonp2}
\end{figure}

\subsection{Multi-ported Torus}
On many torus networks, nodes can transmit simultaneously in all directions. For example, on Fugaku each node is equipped with six TNIs (i.e., NICs), each capable of injecting $54.4$~Gb/s. To fully exploit this capability, on a $D$-dimensional torus we divide the vector used in the collective into $2D$ parts. We then execute $2D$ collectives in parallel, each operating on $n/2D$ bytes, using a separate NIC and transmitting in a different direction.  

For instance, as shown in Fig.~\ref{fig:torus_multiport} for a $2D$ torus, rank~$0$ would normally communicate sequentially in the order east (E) $\rightarrow$ north (N) $\rightarrow$ west (W) $\rightarrow$ south (S). With port parallelization, this order is applied only to $1/4$ of the data. In parallel, another $1/4$ is transmitted in the order N $\rightarrow$ W $\rightarrow$ S $\rightarrow$ E, another $1/4$ with order W $\rightarrow$ S $\rightarrow$ E $\rightarrow$ N, and the last $1/4$ with order S $\rightarrow$ E $\rightarrow$ N $\rightarrow$ W. In this way, all four NICs are active in parallel, thus saturating the available bandwidth. It is worth remarking that on rectangular torus, after some step some of the ports will not be used anymore, since no more communications will happen along those shorter dimensions. 

\begin{figure}[htpb]
  \centering
  \includegraphics[width=\linewidth]{fig/torus/bine_torus_multiport.pdf}
  \caption{Communications happening on a 2D torus when multiported \bine tree algorithms are used.}
  \label{fig:torus_multiport}
\end{figure}

To ensure that each collective uses a different NIC, we implemented the \bine algorithms directly on top of the low-level uTofu library, caching registered remote addresses as we suppose it would likely happen in Fujitsu MPI as well. To simplify the management of several interfaces, each one is managed by a separate thread.

\end{appendix}
\fi

\end{document}